\documentclass[sigconf]{acmart}

\AtBeginDocument{%
  \providecommand\BibTeX{{%
    \normalfont B\kern-0.5em{\scshape i\kern-0.25em b}\kern-0.8em\TeX}}}

\setcopyright{rightsretained}
\acmDOI{}

\acmConference[]{}{}{}
\acmPrice{}
\acmISBN{}

\usepackage[capitalise,nameinlink,noabbrev]{cleveref}
\usepackage{enumitem}
\usepackage{tabularx}
\usepackage{booktabs}
\newcolumntype{H}{>{\setbox0=\hbox\bgroup}c<{\egroup}@{}} 
\usepackage{colortbl}
\usepackage{multirow}
\usepackage{xspace}
\usepackage{float}
\newcommand{\nvm}{PMem\xspace} 
\newcommand{\nvmF}{Persistent Memory\xspace}

\usepackage{csquotes}

\usepackage{subcaption}
\usepackage{tikz}
\usetikzlibrary{calc,shapes}
\newcommand*\circled[1]{\tikz[baseline=(char.base)]{
  \node[shape=ellipse,draw,inner sep=0.75pt] (char) {\parbox{1.6em}{\centering\footnotesize\textbf{#1}}};}}
\newcommand{\tikzmark}[1]{\tikz[remember picture,overlay, baseline=-0.5ex]\node (#1){};}
\newcommand{\tikzoverlay}[2]{\tikz[remember picture,overlay, baseline=-0.5ex]\node at (#1){#2};}

\def\check{\tikz\fill[scale=0.25](0.1,.45) -- (.35,0.1) -- (1,.9) -- (.35,.3) -- cycle;}

\newcommand{\mybox}[1]{\tikz[baseline={([yshift=-.6ex]current bounding box.center)}]{
  \node[shape=rectangle,fill, inner sep=0.5pt, outer sep=0pt,clip, minimum width=1.em, minimum height=1.em,rounded corners=1.pt](char){#1};}}
\renewcommand{\checkmark}{\mybox{\color{white}{\check}}}
\newcommand{\notEval}{\mybox{\color{white}{\check}}}

\graphicspath{{./pics/}}

\begin{document}

\title{Data Structure Primitives on \nvmF: An Evaluation}


\author{Philipp G\"otze}
\email{philipp.goetze@tu-ilmenau.de}
\orcid{0000-0002-5076-5007}
\affiliation{%
  \institution{TU Ilmenau, Germany}
  \vspace{2em}
}

\author{Arun Kumar Tharanatha}
\email{arun-kumar.tharanatha@tu-ilmenau.de}
\affiliation{%
\institution{TU Ilmenau, Germany}
}

\author{Kai-Uwe Sattler}
\email{kus@tu-ilmenau.de}
\orcid{https://orcid.org/0000-0003-1608-7721}
\affiliation{%
\institution{TU Ilmenau, Germany}
}

\begin{abstract}
\nvmF (\nvm), as already available, e.g., with Intel Optane DC Persistent Memory, represents a very promising, next-generation memory solution with a significant impact on database architectures.
Several data structures for this new technology have already been proposed.
However, primarily only complete structures are presented and evaluated.
Thus, the implications of the individual ideas and \nvm features are concealed.
Therefore, in this paper, we disassemble the structures presented so far, identify their underlying design primitives, and assign them to appropriate design goals regarding \nvm.
As a result of our comprehensive experiments on real \nvm hardware, we can reveal the trade-offs of the primitives for various access patterns. This allowed us to pinpoint their best use cases as well as vulnerabilities.
Besides our general insights regarding \nvm-based data structure design, we also discovered new combinations not examined in the literature so far.
\end{abstract}

\maketitle
\thispagestyle{empty}

\section{Introduction}
\label{sect:intro}

Data structures play a crucial role in all data management systems.
Over the past decades, numerous structures have been designed for very different purposes and each design is always a compromise among the three performance trade-offs read, write, and memory amplification \cite{DBLP:conf/edbt/AthanassoulisKM16}.
Furthermore, advances in hardware technology with changing characteristics make designing data structures an ever-lasting challenge.

\nvmF (\nvm) -- also known as non-volatile memory (NVM) or storage-class memory (SCM) -- is one of the most promising trends in hardware development which might have a huge impact on database system architectures in general, but also particularly on data structures.
Intel\textsuperscript{\textregistered} has recently commercialized Optane\texttrademark{} DC Persistent Memory Modules (DCPMMs) based on the 3D XPoint\texttrademark{} technology~\cite{3dxpoint}, on which we focus in this paper.
Characteristics such as byte-addressability, read latency close to DRAM but with a read-write asymmetry, and the inherent persistence open up new opportunities but require also new designs, e.g., to mitigate the read-write asymmetry or to guarantee consistent updates.

Over the last few years, several data structures for \nvm have been proposed trying to address these specifics.
However, the lack of widely available hardware platforms, different benchmarks, and complex designs addressing different aspects make it difficult to compare these approaches and -- more importantly -- identify the most promising \nvm-specific primitives.

Idreos et al.~\cite{DBLP:journals/debu/IdreosZADHKGMQW18} presented the idea of a periodic table of data structures to be able to argue about the design space of data structures.
The work provides a great foundation for a systematic study of data structure designs.
In this paper, we try to support this approach by identifying core primitives of tree-based data structures and evaluate different designs of these primitives on \nvm.
Lersch et al.~\cite{VLDB:lucas} have already extensively evaluated existing B$^+$-Tree designs for \nvm on real hardware.
However, again this was done on the macro level hiding impacts of the separate underlying ideas.
Our contributions are as follows:
\begin{itemize}[leftmargin=*]
  \item From the literature (\textsection\ref{sect:related}), we identify a set of data structure design primitives on \nvm. Furthermore, we generalize the existing approaches for various types of tree-like structures such as B$^+$-Trees, Skip-Lists, Tries, and LSM-Trees (\textsection\ref{sect:primitives}).
  \item We classify these primitives in three \nvm-critical design goals: reducing writes, fine-grained access as well as consistent and durable operations (\textsection\ref{sect:primitives}).
  \item Instead of a black box (or end-to-end) approach, we introduce typical low-level access patterns applicable to the primitives (\textsection\ref{sect:primitives}).
  \item We comprehensively evaluate and report a selection of these access patterns on real \nvm hardware (\textsection\ref{sect:eval}).
  \item We summarize our findings within a performance profile per primitive and formulate general insights (\textsection\ref{sect:eval} and \textsection\ref{sect:conclusion}).
\end{itemize}
To the best of our knowledge, this is the first evaluation considering various data structure types and designs on the micro level with real \nvm hardware.
The goal of our work is to get deep insights into \nvm-optimized design patterns for data structures in databases.

\section{Persistent Memory Properties}
\label{sect:background}


There are several variants of \nvm that use different physical mechanisms to achieve persistence.
PCM~\cite{DBLP:journals/micro/LeeZYZZIMB10} is probably one of the best-known technologies among them.
Intel\textsuperscript{\textregistered} has recently commercialized Optane\texttrademark{} DC Persistent Memory Modules (DCPMMs) based on the 3D XPoint\texttrademark{} technology~\cite{3dxpoint}, which seems to behave similarly to PCM.
What makes \nvm special is the byte-addressability and direct persistence at DRAM speed.
On modern CPU architectures, byte-addressability corresponds to cache-line granularity (typically 64 bytes).
Further interesting features are a higher density and better economic characteristics than DRAM (both in monetary and energy terms) as well as direct load and store semantics.
Another important fact is that the Optane DC devices internally work with cache lines, but a write-combining buffer aggregates writes to 256-byte blocks (cf. \cite{DBLP:conf/damon/RenenVL0K19}).
This is mainly to avoid write-amplification.
In our experiments, we could not identify a performance difference when switching from 64-byte to 256-byte aligned data structures.
Therefore, we assume that it is enough when the data nodes are at least 256~Bytes in size but only are aligned to cache lines.
Another benefit of this buffer is that writes can be faster than reads at low load on the device.
However, that is also why it is hard to measure the real write latency.

\begin{table}[t]
  \small
  \caption{\label{tab:characteristics}Main characteristics of different memory/storage technologies (cf. ~\protect\cite{VLDB:lucas,DBLP:journals/tpds/MittalV16b,DBLP:conf/eurosys/RaoKKLRSJ14, DBLP:conf/damon/RenenVL0K19})}
  \begin{tabularx}{\linewidth}{Xlll}
    \toprule
      & DRAM & Optane DC & NAND Flash \\
    \midrule
    Idle read latency & $80~ns$  & $175~ns$ & $25~\mu{}s$ \\
    Loaded rand. lat. & $120~ns$ & $400~ns$ & $N/A$ \\
    Write latency     & $80~ns$  & $100~ns - 2~\mu{}s$    & $500~\mu{}s$ \\
    Write endurance    & $>10^{15}$ & $N/A$    & $10^{4}-10^{5}$\\
    Density            & $1X$       & $2X-4X$  & $4-8X$\\
  \bottomrule
\end{tabularx}
\end{table}

\cref{tab:characteristics} summarizes some of the characteristics and compares them with those of DRAM and SLC NAND flash.
We remeasured the latencies on our system (see \cref{sect:eval}) using Intel's Memory Latency Checker~\cite{mlc} and Flexible I/O tester~\cite{fio}.
Since we focus on single-threaded experiments, total bandwidth numbers are not relevant for us here.
Similar to flash, \nvm exhibits a read-write asymmetry and lower write endurance than DRAM.
However, we could not find any actual endurance data of the DCPMMs.
When designing new data structures, these properties mean that writes should be minimized using more computing power instead.


The DCPMMs provide two possible operating modes: \texttt{Memory} and \texttt{App Direct} mode.
The Memory mode allows applications to use the DCPMMs as extension to volatile memory, where DRAM acts like a kind of L4 cache.
For that no rewrite of in-memory software is necessary.
However, to fully utilize \nvm and its persistence the App Direct mode must be used.
Therefore, developers have to take care of persistence, failure-atomicity, performance, and so on themselves.
In the remainder of this paper, we exclusively use the latter mode.
On the software level, we used the de facto standard Persistent Memory Development Kit (PMDK)~\cite{pmdk} to get uniform and comparable implementations.
It provides different levels of granularity to manage \nvm including allocations, transactions, object management, etc.

\section{Related Work}
\label{sect:related}


\textbf{Data Structures for \nvm.}
With the properties described in \cref{sect:background} new more fine-grained techniques are enabled when designing \nvm-based data structures.
An overview of these approaches is given in~\cite{DBLP:journals/dbsk/GotzeRLLO18}.
There have already been several publications addressing byte-addressability and write endurance in particular.
One of the first approaches of Venkataraman et al.~\cite{DBLP:conf/fast/VenkataramanTRC11} proposes a single-level storage hierarchy and general ideas for consistent and durable data structures.
They mainly focus on B$^+$-Trees and use versioning, atomics, and shadowing to guarantee atomicity.
This is also addressed by Chen et al.~\cite{DBLP:journals/pvldb/ChenJ15} who exploit indirection and propose to keep nodes unsorted to save writes.
Additionally, they compare the approaches and effects when adding certain features such as bitmaps.
Yang et al.~\cite{DBLP:conf/fast/YangWCWYH15} propose selective consistency, i.e., enforcing consistency of leaf nodes and relaxing it for inner nodes.
Here, too, leaf nodes are kept unsorted and new keys are just appended.
In~\cite{DBLP:conf/sigmod/OukidLNWL16} Oukid et al. present a hybrid solution where leaf nodes remain in the persistent layer but inner nodes are placed in DRAM.
This allows a much faster traversal of the upper levels but requires recovery measures to rebuild it in case of failure.
Another crucial part is the use of fingerprints to reduce the number of keys probed. 
HiKV~\cite{DBLP:conf/usenix/XiaJXS17} takes a similar path and places the B$^+$-Tree in DRAM and holds only hash partitions persistent.
This avoids costly structure reorganizations on \nvm.
The B$^P$-Tree~\cite{DBLP:journals/tkde/HuLNST14} buffers changes in DRAM and, when full, merges them into \nvm.
They collect information to predict future accesses to pre-allocate nodes and reduces writes caused by splits or merges.
In~\cite{DBLP:journals/tos/KimSKN18} the authors propose cache-line-sized nodes combined with differential encoding to reduce the number of cache line flushes.
The BzTree~\cite{DBLP:journals/pvldb/ArulrajLML18} is a high-performance latch-free B-Tree using a persistent multi-word compare-and-swap (PMwCAS~\cite{DBLP:conf/icde/WangLL18}) operation to provide failure atomicity.
In~\cite{VLDB:lucas} some of these trees are already evaluated on real \nvm hardware.
But again the complete trees were compared, instead of the individual primitives.
As a result, e.g., the wB$^+$-Tree~\cite{DBLP:journals/pvldb/ChenJ15} always performs poorly as the inner nodes are persistent, leading to a costly traversal.
Furthermore, their evaluation is limited to B$^+$-Trees.

Recalling the properties of \nvm, write-optimized data structures such as the LSM-Tree are a promising option.
Many modern key-value stores such as RocksDB~\cite{rocksdb} or Cassandra~\cite{cassandra} are based on this concept.
There are already first approaches to adapt this concept for \nvm~\cite{DBLP:conf/usenix/KannanBGAA18,DBLP:conf/damon/LerschOLS17,nvmrocks}.
Furthermore, prefix trees (tries) like ART were already adapted to \nvm~\cite{art_pm} as well as some write optimized versions of it~\cite{DBLP:conf/fast/LeeLSNN17}.
There are also first publications in the field of hash tables~\cite{DBLP:conf/sosp/DebnathHKKU15,DBLP:conf/vldb/SchwalbDUP15,DBLP:conf/osdi/Zuo0W18}.
So far, only~\cite{DBLP:conf/icde/Gotze0S18} has targeted analytical processing, whose idea is based on clustering and unsorted blocks.
The special feature is the three-level architecture and the ability to efficiently query any attribute besides the key.

The approaches mentioned above have so far been mainly evaluated for operator or end-to-end performance.
However, this hides important details and trade-offs of the underlying design primitives on which we focus in this paper.

\begin{table*}[t]
  \small
  \caption{\label{tab:primitives}Design primitives and micro-operations for \nvm-aware trees
(\protect\checkmark~$\rightarrow$ applicable).}
  \vspace{-1em}
  \begin{tabularx}{\linewidth}{p{1.7cm}X|ccc|cccc|ccc|c}
    \toprule

    & &
    \multicolumn{3}{c|}{\cellcolor[gray]{0.9}Read-based} & \multicolumn{4}{c|}{\cellcolor[gray]{0.9}Insert-based} & \multicolumn{3}{c|}{\cellcolor[gray]{0.9}Erase-based} & \multirow[b]{2}{*}{\rotatebox[origin=r]{70}{Recovery}}\\
    \cmidrule{3-12}
    \multicolumn{2}{l|}{\raisebox{4em}{\scalebox{1.75}[1.75]{\nvm-aware Trees}}}
     &
    \rotatebox[origin=l]{70}{Node Search} & \rotatebox[origin=l]{70}{Tree Traverse} & \rotatebox[origin=l]{70}{Tree Iterate} &
    \rotatebox[origin=l]{70}{Insert in Node} & \rotatebox[origin=l]{70}{Node Split} & \rotatebox[origin=l]{70}{Move Node} & \rotatebox[origin=l]{70}{Merge Level} &
    \rotatebox[origin=l]{70}{Erase f. Node} & \rotatebox[origin=l]{70}{Balance Nodes}  & \rotatebox[origin=l]{70}{Merge Nodes}
    & \\

    \midrule
    \multirow[t]{6}{\hsize}{\cellcolor[gray]{0.9}\textbf{D}{\footnotesize esign}\textbf{G}{\footnotesize oal}\textbf{1}\newline(reduce writes)}
     & sorted~\cite{DBLP:journals/pvldb/ArulrajLML18,DBLP:journals/tkde/HuLNST14,DBLP:conf/fast/VenkataramanTRC11} & \tikzmark{p1} & & \tikzmark{p2} & \checkmark & \checkmark & \checkmark &  & \checkmark & \checkmark & \checkmark & \\
     & unsorted~\cite{DBLP:journals/pvldb/ArulrajLML18,DBLP:journals/pvldb/ChenJ15,DBLP:journals/tkde/HuLNST14,DBLP:journals/tos/KimSKN18,DBLP:conf/sigmod/OukidLNWL16,DBLP:conf/fast/YangWCWYH15} &  &  &   & \checkmark & \checkmark & \checkmark &  & \checkmark & \checkmark & \checkmark & \\
     & bitmaps~\cite{DBLP:journals/pvldb/ChenJ15,DBLP:journals/tos/KimSKN18,DBLP:conf/sigmod/OukidLNWL16} &  &  &  & \checkmark & \checkmark &  &  & \checkmark & \checkmark & \checkmark & \\
     & indirection~\cite{DBLP:journals/pvldb/ChenJ15} &  & &   & \checkmark & \checkmark & & & \checkmark & \checkmark & \checkmark & \\
     & hashing~\cite{DBLP:conf/sigmod/OukidLNWL16, DBLP:conf/usenix/XiaJXS17} & & &  & \checkmark & \checkmark &  &  & \checkmark & \checkmark & \checkmark & \\
     & split move~\cite{DBLP:journals/pvldb/ArulrajLML18,DBLP:journals/pvldb/ChenJ15,DBLP:journals/tkde/HuLNST14,DBLP:journals/tos/KimSKN18,DBLP:conf/fast/VenkataramanTRC11,DBLP:conf/fast/YangWCWYH15} & \tikzmark{p3} &   &  \tikzmark{p4}  & & \checkmark &  &  &  &  & & \\
    & 2-way merge & & & &  &  &   &\checkmark   &   &    &   \\
    & K-way merge & & & &  &  &   &\checkmark    &   &    &   \\
    & placement~\cite{DBLP:journals/tkde/HuLNST14,DBLP:conf/sigmod/OukidLNWL16,DBLP:conf/usenix/XiaJXS17,DBLP:conf/fast/YangWCWYH15} & \tikzmark{p5} \checkmark  & \checkmark & \tikzmark{p6} & \checkmark & \checkmark & \checkmark & \checkmark & \checkmark & \checkmark & \checkmark & \notEval\\
    \midrule

    \multirow[t]{4}{\hsize}{\cellcolor[gray]{0.9}\textbf{D}{\footnotesize esign}\textbf{G}{\footnotesize oal}\textbf{2}\newline(fine-grained access)}
     & linear search~\cite{DBLP:journals/pvldb/ArulrajLML18,DBLP:journals/tkde/HuLNST14,DBLP:journals/tos/KimSKN18,DBLP:conf/fast/YangWCWYH15} & \checkmark & \notEval  & \checkmark & \checkmark & \checkmark & &  & \checkmark & \checkmark & \checkmark & \notEval\\
     & search with bit check~\cite{DBLP:journals/pvldb/ChenJ15,DBLP:journals/tos/KimSKN18,DBLP:conf/sigmod/OukidLNWL16} &  \checkmark & \notEval & \checkmark & \checkmark & \checkmark & \ & & \checkmark & \checkmark & \checkmark & \notEval \\
     & search with hash probing~\cite{DBLP:conf/sigmod/OukidLNWL16} & \checkmark &  \tikzmark{p7} &  & \checkmark & \checkmark & & & \checkmark & \checkmark & \checkmark & \notEval\\
     & binary search~\cite{DBLP:journals/pvldb/ArulrajLML18,DBLP:journals/tkde/HuLNST14,DBLP:conf/fast/VenkataramanTRC11,DBLP:conf/fast/YangWCWYH15} & \checkmark & \notEval &  & \checkmark & \checkmark & & & \checkmark & \checkmark & \checkmark & \notEval\\
     & search with indirection~\cite{DBLP:journals/pvldb/ChenJ15} & \checkmark & \notEval & \checkmark & \checkmark & \checkmark & & & \checkmark & \checkmark & \checkmark & \notEval\\
     & split copy~\cite{DBLP:conf/sigmod/OukidLNWL16} & \tikzmark{p10}  &  & \tikzmark{p11}  &  & \checkmark & & & & & & \notEval\\
     & cache sensitive~\cite{DBLP:journals/pvldb/ArulrajLML18,DBLP:journals/pvldb/ChenJ15,DBLP:journals/tos/KimSKN18,DBLP:conf/sigmod/OukidLNWL16,DBLP:conf/fast/YangWCWYH15} & \checkmark & \checkmark & \checkmark & \checkmark & \checkmark & \checkmark & \checkmark & \checkmark & \checkmark & \checkmark & \notEval\\

    \midrule
    \multirow[t]{3}{\hsize}{\cellcolor[gray]{0.9}\textbf{D}{\footnotesize esign}\textbf{G}{\footnotesize oal}\textbf{3}\newline(failure atomicity)}
     & PMDK Transactions~\cite{DBLP:conf/icde/Gotze0S18,pmdk} & \tikzmark{p8} & & & \checkmark & \checkmark & \checkmark & \checkmark & \checkmark & \checkmark & \checkmark & \notEval \\
     & PMwCAS~\cite{DBLP:journals/pvldb/ArulrajLML18,DBLP:conf/icde/WangLL18} & & & &  \notEval &  \notEval &  \notEval &  \notEval &  \notEval &  \notEval  &  \notEval & \notEval \\
     & individually~\cite{DBLP:journals/pvldb/ChenJ15,DBLP:journals/tos/KimSKN18,DBLP:conf/sigmod/OukidLNWL16,DBLP:conf/fast/VenkataramanTRC11,DBLP:conf/usenix/XiaJXS17,DBLP:conf/fast/YangWCWYH15} &   &   & \tikzmark{p9}  & \checkmark & \checkmark & \checkmark & \checkmark & \checkmark & \checkmark & \checkmark & \notEval\\
     \midrule[\heavyrulewidth]
     \multicolumn{2}{c|}{Evaluation reported in Experiment No.}
      & \textbf{E1} & \textbf{E2} & \textbf{E3} & \textbf{E4} & \textbf{E5} & \textbf{E6} & \textbf{E7} & \textbf{E8} & \textbf{E9} & \textbf{E10} &  \\

   \bottomrule
   \tikzoverlay{4.65,8.05}{\footnotesize \textit{design primitives}}
   \tikz[remember picture,overlay]\draw[shorten <=-0mm, shorten >=-0mm](4.7,9.0)--(5.95,8);
   \tikzoverlay{5.75,9.05}{\footnotesize \rotatebox[origin=l]{90}{\textit{micro-operations}}}
  \end{tabularx}
  \vspace{-2.em}
\end{table*}

\medskip

\textbf{Evaluating Data Structure Designs.}
Several approaches analyze access patterns and the hardware profile to pick appropriate data structures and implementations as well as hardware placement.
Similar to us, some of these also subdivide the data structures into primitives.
The Data Calculator~\cite{DBLP:conf/sigmod/IdreosZHKG18} and the periodic table of data structures~\cite{DBLP:journals/debu/IdreosZADHKGMQW18}, for example, discuss a novel approach which interprets data structures as an assembly of first principles.
The authors combine analytical models, benchmarks, and machine learning to gain insights into the impact of these fundamental primitives.
Their engine takes a high-level specification of a data structure assembled from the primitives and predicts the performance for a given workload and hardware profile.
The so-called operation and cost synthesizer learns a basic set of cost models for different access patterns and synthesizes the cost for more complex operations.
These cost models are trained by micro-benchmarks and thus strongly dependent on these.
It is indicated that the benchmarks must be entered manually by the user when new patterns or hardware is added.
This is where the micro-benchmarks presented in this paper can tie in very well.


\section{Design Primitives}
\label{sect:primitives}

\textbf{Design Goals.}
Using the \nvm properties and related work, we identify three design goals.
The first design goal is the reduction of writes (\textbf{DG1}), which is based on the write endurance and read-write asymmetry.
Due to the byte-addressability, much finer-grained accesses are possible that should be exploited (\textbf{DG2}).
The direct load and store semantics further enable zero-copy memory mapping and, thus, new opportunities to ensure consistency and durability, e.g., by atomic primitives (\textbf{DG3}).

In the following, we start by classifying typical data structures found in DBMSs and give an insight into their huge design space.
Subsequently, we extract the design primitives with focus on tree-based structures from the literature and connect them with our defined design goals.

\begin{figure}[b]
  \centering
  \includegraphics[width=\linewidth]{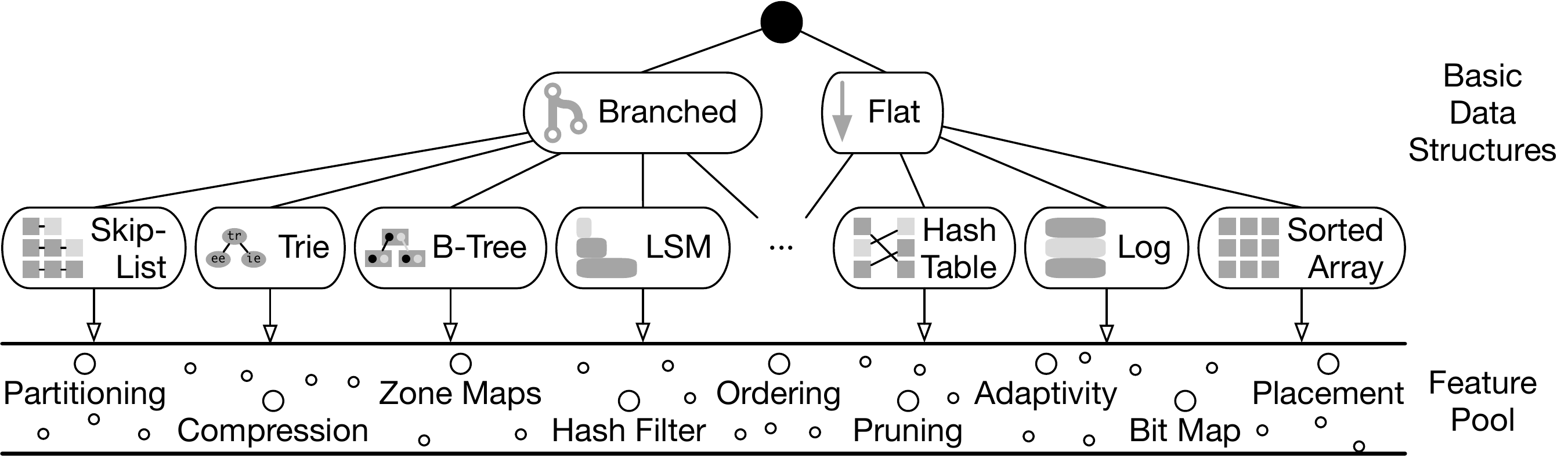}
  \caption{\label{fig:classes}Classification of typical data and index structures in database management systems}
\end{figure}

\medskip
\textbf{Glimpse into the Design Space.}
In \cref{fig:classes}, the typical data and index structures used within a DBMS are summarized.
Each of these data structures is more suitable for certain scenarios and access patterns than others.
As described in~\cite{DBLP:journals/debu/IdreosZADHKGMQW18} there is a general trade-off between read, write, and space optimized designs.
Accordingly, their performance depends on both the workload running on them and the underlying hardware.
Furthermore, these basic structures can also be extended by features or combined with other structures in order to meet the given requirements.
By adding features, sometimes also new access primitives become possible applicable to them.

Looking at the illustration and the related work above, it becomes clear that the design space is huge and there are still thousands of variants that have not been studied yet, in particular with regard to \nvm.
We have primarily worked with B$^+$-Trees, LSM-Trees and single node structures since these already cover a vast part of the design space.

The question we want to answer is what impact certain design primitives have in which scenarios in the presence of \nvm.
The goal is to reveal their trade-offs for facilitating design decisions.
This must be done in the form of white-box testing to avoid side effects in the measurements.
As a result, we envisage a profile per design primitive, from which performance and memory impacts can be derived for each kind of access pattern.

\medskip
\textbf{Definitions.}
Similar to~\cite{DBLP:journals/debu/IdreosZADHKGMQW18}, we define a \emph{design primitive} in this context as an indivisible layout or access concept.
To achieve the goal mentioned above, it is necessary to break down the possible primitives taking into account the properties of \nvm.
For that, we study the approaches as described in~\cref{sect:related} and assign the ideas to the corresponding design goals.
Furthermore, we consider existing micro-operations for trees/nodes and have set these in relation to the derived primitives.
In this context, a \emph{micro-operation} describes a low-level access pattern whose result is independent of the chosen primitive(s).
The typical macro-operation like get, insert, update, delete, and scan can be implemented by combining such micro-operations.
Therefore, we classify them in read-, insert- and erase-based as well as recovery operations.
\cref{tab:primitives} shows our results.
For design primitives that are not applicable or relevant for certain operations, we left the cells empty.

\medskip
\textbf{Micro-Operations.}
The great advantage of considering micro-operations is the disclosure of bottlenecks and optimization potential, which would be concealed at the macro level.
For instance, inserts for hybrid DRAM/\nvm solutions are always faster than completely persistent ones almost independent of the node layout.
Therefore, it is important to keep both access patterns and the design space concise.
For the read category, the first micro-operation is the search for a key within a node (lookup).
To get the target node, there are usually two types of traversing the tree, namely vertical (tree traverse from top) and horizontal (tree iterate from lowest left).
The macro-operations get and scan can be built by combining lookups and traversals.
Next, there are insert-based operations like placing records (e.g., key-value pairs) into a node.
This can lead to splits, which require the allocation of new nodes.
An insert or update macro-operation would need the micro-operations lookup, traversal, insert, and split.
For hybrid structures (such as the  B$^P$- or LSM-Trees) a common operation is also the movement or migration from DRAM to \nvm.
Furthermore, a compaction or merge of multiple nodes (in a level) into a larger node can be necessary.
Erasing an entry from a node is a micro-operation downsizing the tree.
This may cause an underflow which can be resolved by balancing or merging with another node.
The typical delete macro-operation consists of search, traversal, erase, balance and merge.
The last class is recovery.
It mainly consists of the operations of the read category combined with the recreation of volatile DRAM structures.
Persisting operations depend on the primitives of DG3.

\begin{figure}[t]
\begin{subfigure}[b]{\linewidth}
  \includegraphics[height=2.85em]{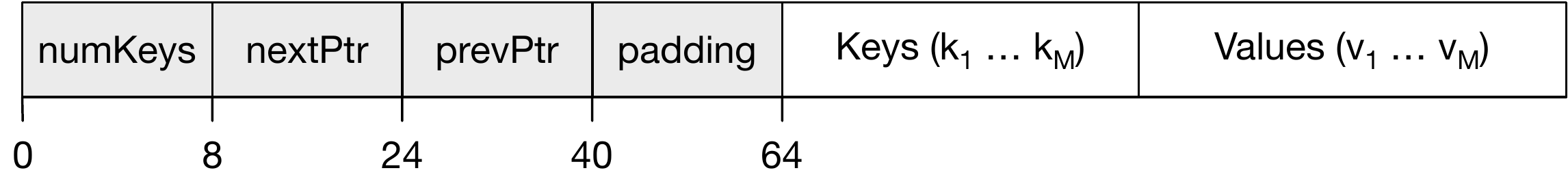}
  \vspace{-.5em}
  \caption{Sorted/Unsorted data nodes.}
  \vspace{1em}
\end{subfigure}
\begin{subfigure}[b]{\linewidth}
  \includegraphics[height=2.85em]{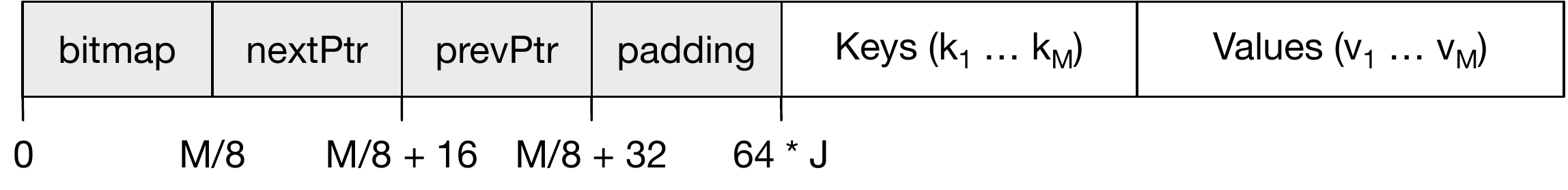}
  \vspace{-.5em}
  \caption{Bitmap-only data nodes.}
  \vspace{1em}
\end{subfigure}
\begin{subfigure}[b]{\linewidth}
  \includegraphics[height=2.85em]{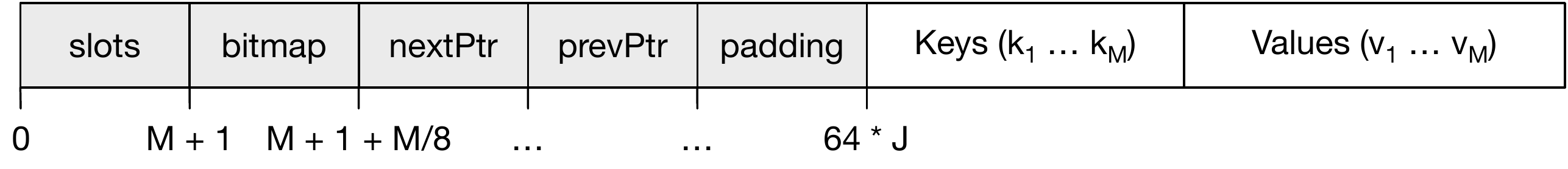}
  \vspace{-.5em}
  \caption{Indirection data nodes.}
  \vspace{1em}
\end{subfigure}
\begin{subfigure}[b]{\linewidth}
  \includegraphics[height=2.85em]{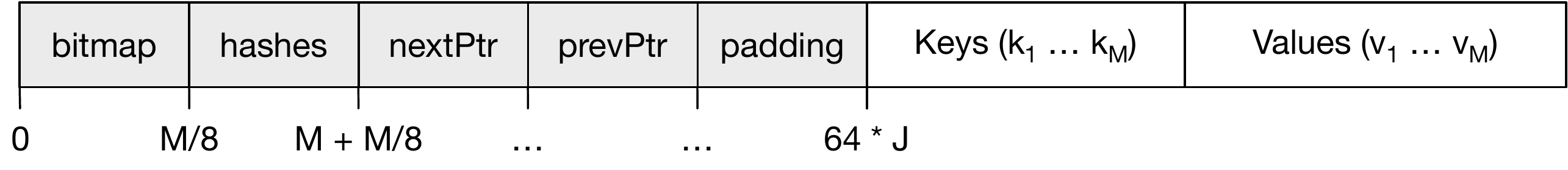}
  \vspace{-.5em}
  \caption{Hashing data nodes.}
\end{subfigure}
\vspace{-1.5em}
\caption{\label{fig:leafStructures}Different data node layouts, where: M - number of entries and $J\in\mathbb{N}_{>0}$. }
\end{figure}

\medskip
\textbf{Primitives.}
We now briefly describe a set of found primitives.
For the first design goal - reducing writes - the node layout (which applies to a large variety of tree-like structures) was reconsidered and the main consensus was to leave data nodes unsorted.
To keep the access fast indirection, hashing, and bitmaps, as well as combinations of these, were used.
For this, appropriate auxiliary structures are added at the start of each node.
In \cref{fig:leafStructures}, we compare the data node layouts of these ideas, how we have reimplemented them for the evaluation (cf. \cref{sect:eval}).

All of them align the search structure in the beginning and the keys to cache lines to have a fair comparison.
For the sorted and unsorted case the metadata always costs only one cache line whereas the other layouts can cover multiple cache lines depending on the number of elements.
To further save writes, the node placement was adapted leading to selective persistence by placing some parts (e.g., inner nodes) in DRAM.
Depending on the node layout there are different access primitives (DG2).
A simple linear search over the keys is always possible.
If the entries are sorted, a binary search is applicable.
Both algorithms can also be modified with more fine-granular access using the cache-line-sized auxiliary structures.
Additionally, other algorithms such as interpolation or exponential search are conceivable.
When splitting a node, as typical in B-Trees and Skip-Lists, we identified two approaches.
The basic algorithms move all keys greater than the split key to the new node.
This can also be done by creating two new nodes.
An alternative when using a bitmap is to copy the full node, reset the greater keys in the bitmap, and finally store the inverted bitmap in the new node.
The first variant will trigger less writes, but the second variant could be faster by exploiting the fine-grained access.
Regarding failure atomicity (DG3), PMDK provides general-purpose transactions that can be placed around the algorithms and allocations.
We will go into more detail below.
Alternatively, PMwCAS could be used, which provides compare-and-swap operations for ranges bigger than eight bytes.
Another method is to individually persist the data using flush and fence instructions.

\begin{figure}[t]
	\centering
	\begin{subfigure}[b]{0.33\textwidth}
		\centering
	 \includegraphics[width=\linewidth]{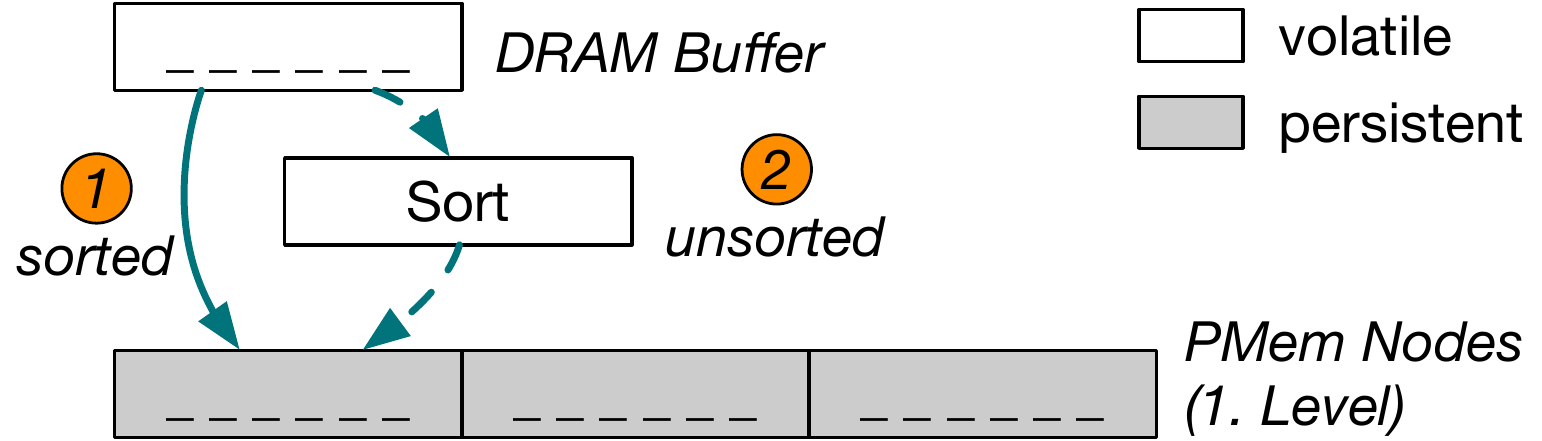}
		\caption{\label{fig:lsm_c0push}Move DRAM data to a \nvm node.}
    \vspace{1em}
	\end{subfigure}

	\begin{subfigure}[b]{0.38\textwidth}
		\centering
	 \includegraphics[width=\linewidth]{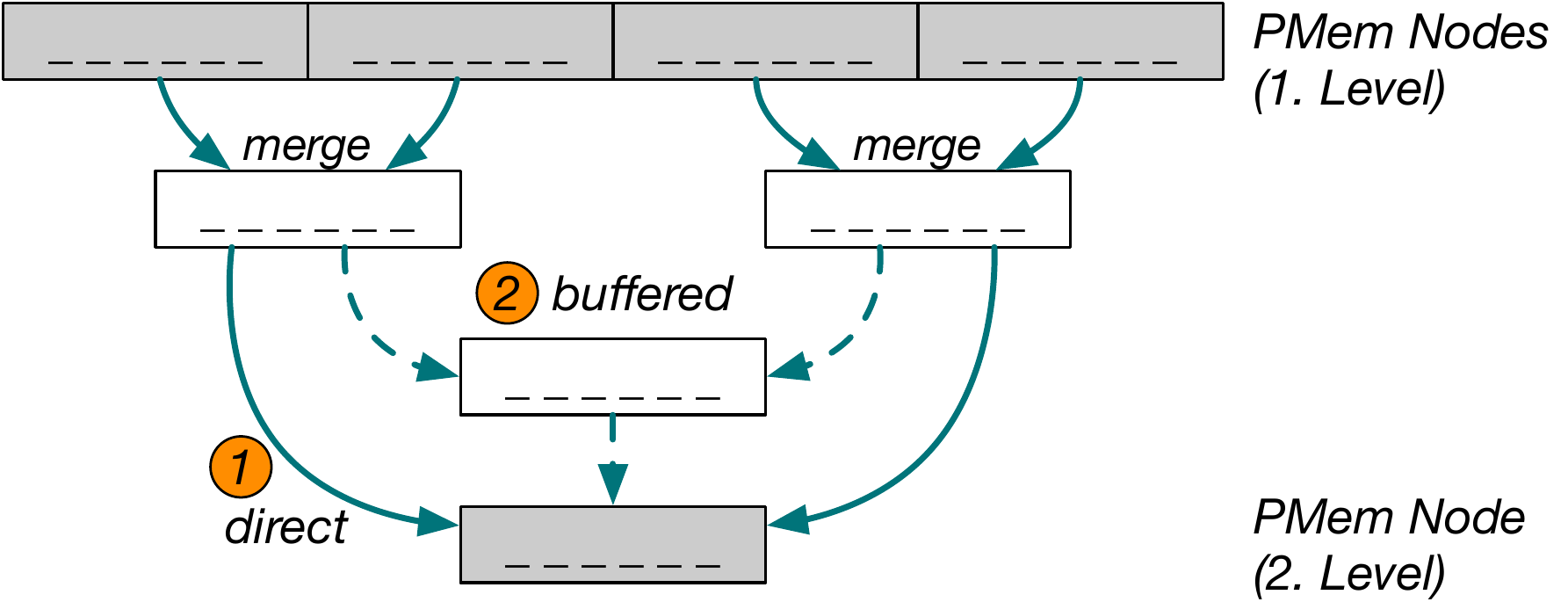}
		\caption{\label{fig:lsm_2wayMerge} 2-way merge into a \nvm node.}
    \vspace{1em}
	\end{subfigure}

		\begin{subfigure}[b]{0.38\textwidth}
		\centering
	 \includegraphics[width=\linewidth]{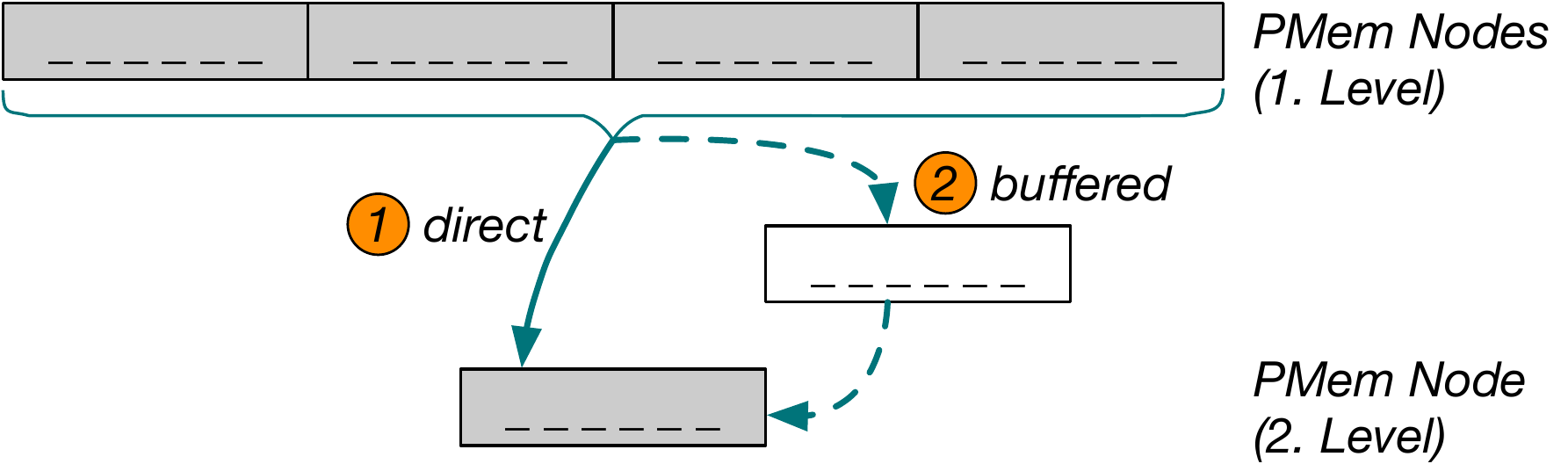}
	\caption{\label{fig:lsm_KwayMerge} K-way merge into a \nvm node.}
	\end{subfigure}
  \vspace{-1em}
	\caption{\label{fig:Lsm_primitives}Design primitives for Move node and Merge level.}
\end{figure}

\cref{fig:lsm_c0push} illustrates the design primitives considered for the \textit{Move Node} operation, as typical for LSM-Trees.
The initial data is stored in a DRAM buffer and later moved to a free persistent node.
We consider a scenario where the data in the persistent nodes is always sorted, whereas the DRAM data can be (1) sorted or (2) unsorted.
If sorted, then, DRAM data is just copied to the \nvm node, else it must be sorted before a copy operation.
In the former case, insertions into DRAM would be costlier and the penalty would be higher for a bigger DRAM node size.
Alternatively, an unordered structure could be used whose insertion costs are comparatively less dependent on DRAM node size.
However, a sort operation is needed during the data movement from DRAM to a persistent node.

\cref{fig:lsm_2wayMerge} and  \cref{fig:lsm_KwayMerge} show the design primitives used for \textit{Merge Level}.
While the design space for merge algorithms is exhaustive, we use two common approaches: 2-way and K-way merge. 
When all the nodes in a level are filled, the sorted data in these nodes are merged to a free persistent node at the next higher level.
The final result of a 2-way or K-way merge could be written using the following two approaches.
(1) Directly perform merge in \nvm.
(2) Merge to a DRAM buffer and copy the result to \nvm.
The former approach could result in a performance benefit when the inserted keys are unique, whereas the latter approach could be a better choice if there are many updates (i.e., duplicate keys).

\begin{figure}[b]
	\centering
	\includegraphics[width=0.55\linewidth]{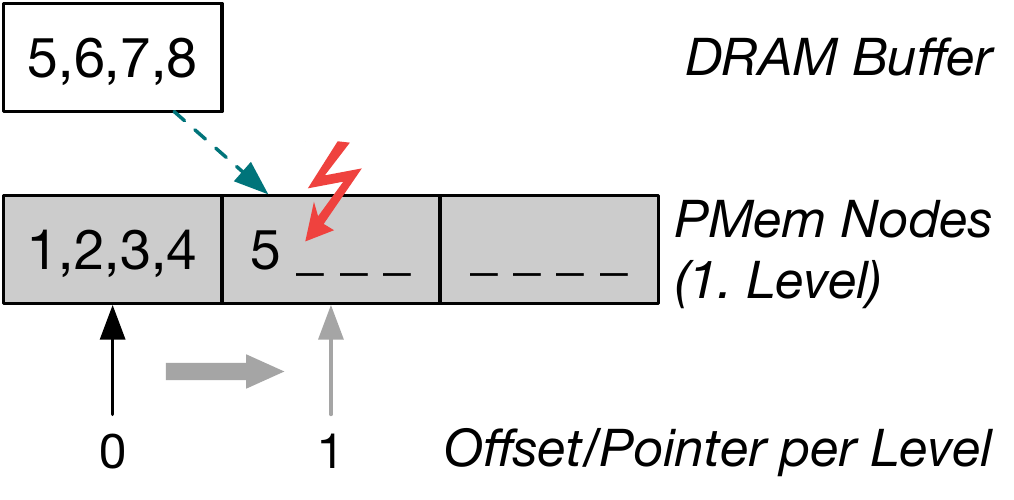}
	\caption{\label{fig:LSM_transactions}Alternative for realizing failure atomicity.\newline (Use case: \nvm-aware LSM-Tree).}
\end{figure}

Using PMDK transactions for failure atomicity (FA), will induce a noticeable performance penalty.
\cref{fig:LSM_transactions} shows a different way of realizing FA for \textit{Move Node} and \textit{Merge Level}.
An array of pointers/offsets is used to keep track of nodes that contain valid data and point to the next free node as shown in \cref{fig:LSM_transactions}.
After moving data from one level to the next, the offset of this level is incremented. 
Suppose a failure occurs during a write to node 1; an undo operation on this node or a portion of it is redundant since the pointer remains at the previous position indicating node 1 is still free. 
Therefore, it is sufficient to only add the pointer into the PMDK transaction. 
As a consequence, the performance penalty of PMDK transactions could be greatly minimized.
We term this as \textit{Individual failure atomicity}.
However, the data must be flushed before the pointer with the help of fences.
Furthermore, on x86 architectures, 8-byte aligned writes are failure atomic.
Hence, by limiting the size of the offsets to 8 bytes and properly aligning then, we can completely avoid transactions.
We term this as \textit{No FA}.
Regarding \cref{tab:primitives}, these mechanisms basically fall under DG3, whereby \textit{Individual FA} and \textit{No FA} also fit into DG1.

\medskip
\textbf{Extendability.}
\Cref{tab:primitives} is only an excerpt and can be extended by more primitives and micro-operations.
Further aspects would be, e.g., hardware utilization, concurrency, and more in-depth failure atomicity.
We excluded concurrency control as it does not fit into our micro consideration and is also sufficiently complex to merit a separate paper.
Regarding hardware utilization, we already applied cache-line alignment for all nodes and auxiliary structures.

\medskip
\textbf{Metrics.}
The task now is to check and evaluate the applicable options in the table with the help of micro-benchmarks.
We have already pursued this task to a certain extent.
However, before delving into the experiments, we must first define relevant metrics.
Since we are on the micro level, throughput does not provide usable values at this point.
Therefore, we mainly report the average latency per operation.
Moreover, hardware-specific measures can be studied such as cache misses, flushes, instructions per cycle, or the number of reads and writes. From our point of view, the number of persist operations or written bytes are crucial factors due to the read-write asymmetry.
Apart from the performance indicators, memory consumption is of interest, as \nvm is less dense than disks.

\section{Experiments}
\label{sect:eval}

In our experiments, we focus on the micro-operations on tree-like data structures as introduced in the previous section.
From the primitives described above, we picked for the node layout: sorted, unsorted, indirection + bitmap ("indirection"), hash-probing + bitmap ("hashing") and bitmap only, in most of the experiments.
For the access primitives, binary search with and without using indirection as well as linear search with and without using hashing and bitmaps are tested.
We re-implemented the approaches from the literature focusing on the corresponding primitive(s).
More details are given at each experiment.
The aim and contribution are to evaluate the design primitives independently of their original context and to compare their strengths and costs.
This should reveal a performance profile for each primitive sketched at the end of this section.

\begin{table}[b]
  \small
  \caption{\label{tab:setup}Experimental setup.}
\begin{tabularx}{\linewidth}{lX}
  \toprule
  \textsc{Processor} & 2 Intel\textsuperscript{\textregistered}~Xeon\textsuperscript{\textregistered}~Gold 5215, 10 cores / 20 threads each, max. 3.4~GHz \\\midrule
  \textsc{Caches} & 32~KB L1d, 32~KB L1i, 1024~KB L2, 13.75~MB LLC \\\midrule
  \textsc{Memory} & 2$\times$6$\times$32 GB DDR4 (2666 MT/s), \newline 
                    2$\times$6$\times$128 GB Intel\textsuperscript{\textregistered} Optane\texttrademark~DCPMM 
                    \\\midrule
  \textsc{OS \& Compiler} & CentOS 7.8, Linux 5.6.11 kernel, cmake 3.15.3,\newline ICC~19.1.0.166 (-O3) \\\bottomrule
\end{tabularx}
\end{table}

\begin{figure*}
  \centering
  \includegraphics[width=1.\linewidth]{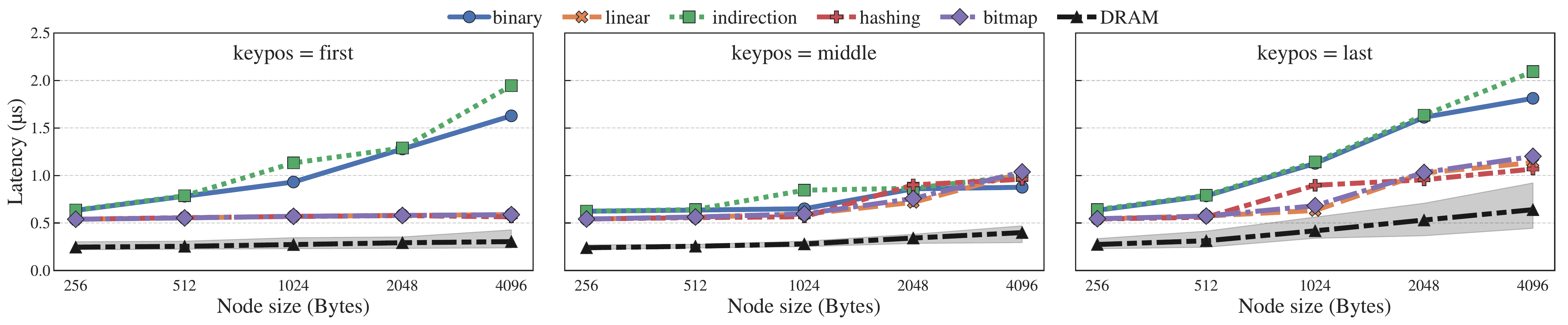}
  \caption{\label{fig:leafsearch}Searching for a key within a node (E1).}
\end{figure*}

\subsection{Experimental Setup}
For our experiments, we used a dual-socket Intel Xeon Gold 5215 server as outlined in \cref{tab:setup}.
Each socket comes with 6 DCPMMs, which we grouped into one region and namespace. 
The operating mode of the modules is set to \texttt{AppDirect} allowing direct access to the devices.
On the \nvm DIMMs, we created an \texttt{ext4} file system and mounted it with the \texttt{DAX} option to enable direct loads and stores bypassing the OS cache.
To avoid NUMA effects, all experiments are controlled to allocate resources (memory, persistent memory, and cores) only from the same socket.

We are using PMDK~\cite{pmdk} for the implemented data structures to guarantee failure atomicity.
Alternatively, PMwCAS~\cite{DBLP:conf/icde/WangLL18} could be used, which provides \texttt{CAS} operations even for structures bigger than 8~bytes.
Another method could be to manually place persist, flush, or fence instructions, which is also possible with PMDK.
Since the transactions of PMDK had so much overhead hiding the impact of the approaches in our implementations, we decided to mainly report the results for manually persisting the modified data.

Unless stated otherwise, we used fixed-size keys and values being 8-byte integers and 16-byte tuples ($<\negthickspace int, int, double\negthickspace>$), respectively, in all our experiments.
The size of the values, therefore, corresponds to the size of a persistent pointer (e.g., to the actual payload).
Keys, values, and children pointer were stored in separate arrays within the nodes for better locality benefits when iterating through the keys.
In addition, all nodes as well as their inner key arrays are aligned to cache lines.
The fill ratio of the trees was always 100\% to make optimal use of memory.
Since, in most experiments, we access predefined positions, the fill ratio is not too important.
When the node size is varied the various implementations often result in a different number of actual elements due to their node layout.
To primarily measure \nvm and not cache performance, we created a collection of nodes (or trees) that is a multiple of the size of the LLC.
Each iteration randomly accesses one of these instances to reduce prefetching and caching of other instances as much as possible.
When referring to a cached case below, we mean that each iteration always accesses the same instance.
Every data point in our plots is supported by several thousand iterations.
Our implementations can be examined via our public repository.\footnote{\scriptsize \url{https://github.com/dbis-ilm/PMem_DS}}

\subsection{Read Operations}
\medskip
\textbf{Node Search (E1).}
In our first experiment, we study the performance for searching a key position within a node.
This type of operation is crucial to nearly every macro operation including get, update, and delete.
We varied both the node sizes as well as the position of the requested key and tested on various node layouts combined with their corresponding access primitive.
The expectation is that the approaches using binary search are faster, except for accesses to front elements.
\cref{fig:leafsearch} shows our outcome including a baseline that aggregates the results for all variants on DRAM.

We observe that our expectations have not been met in this setup.
Only if the key is in the middle binary and linear approaches show a similar performance.
The indirection approach is a little bit slower than the direct binary search but will cost much less writes for inserts and deletes - which we will consider later.
The disadvantage of indirect binary search is that it needs to jump back and forth from search structure to key array.
Of the linear approaches, hashing is usually the best since all comparisons are first done in the front cache line(s) and only if a hash matches, the actual key is checked.
It means that on average, the fewest cache lines have to be loaded from \nvm.
However, it should be noted that for DRAM the binary search is better in the middle and, for a cached setup, also in the back access areas.
Furthermore, the unsorted structures are not suitable for inner nodes, since the search is not based on equality but a key range.
This is certainly relevant to hybrid structures.
It is also notable that the lines for indirection and hashing are sometimes jumpy.
This is due to the changing size of the front search structure depending on the node capacity.
For 1~KiB and 2~KiB, the search structure consumes two cache lines and for 4~KiB it needs three cache lines.
However, in contrast to indirection, hashing usually reads less of these cache lines.
Thus, indirection and hashing (and partly also the bitmap) require more memory.

\begin{table}[b]
  \small
  \caption{\label{tab:keysLeafNode}Calculated number of records per node (r/n) and memory consumption of a node chain (50M records) for a given node size.}
  \begin{tabularx}{\linewidth}{Xrrrrr|r}
    \toprule
       & \textsc{256~B} & \textsc{512~B} & \textsc{1~KiB} & \textsc{2~KiB} & \textsc{4~KiB} & \textsc{Unit}\\
    \midrule
    \multirow[t]{2}{\hsize}{\textsc{Base}}
    &    9 &   19 &   41 &   83 &  169 & r/n \\
    & 1,32 & 1,25 & 1,16 & 1,15 & 1,13 & GB\\
    \multirow[t]{2}{\hsize}{\textsc{Aligned}}
    &    8 &    18 &   40 &   82 &  168 & r/n \\
    & 1,49 &  1,32 & 1,19 & 1,16 & 1,14 & GB\\
    \multirow[t]{2}{\hsize}{\textsc{w/ search\\ structure}}
    &    8 &   18 &   37 &   79 &  160 & r/n \\
    & 1,49 & 1,32 & 1,25 & 1,22 & 1,19 & GB\\
    \midrule
    \textsc{Overhead Aligned}        & +13\% & +6\% & +3\% & +1\% & +1\% & \\
    \textsc{Overhead Search Structure} & +13\% & +6\% & +8\% & +6\% & +5\% & \\

  \bottomrule
\end{tabularx}
\end{table}

Talking about memory, \cref{tab:keysLeafNode} shows the actual number of entries that can be stored for a given node size.
Without a search structure -~as for basic binary and linear search~- more records can be placed in a node.
For indirection and hashing each entry requires an additional bit for the bitmap and an additional byte for the slot or hash array.
For a fairer comparison, we also aligned the approaches without a search structure so that the counter of entries and the sibling pointers are placed in the first cache line.
The resulting size adjustment can also be found in the table.
Hence, all variants have their actual data cache-aligned as already mentioned above.
It becomes visible that smaller node sizes generally lead to a larger overhead to the \nvm consumption.
In addition, this also results in a longer traversing path.
This of course highly depends on the size of the keys and values.
Apart from the higher memory footprint, hashing is the best choice for searching a persistent node.
Nodes that are in DRAM or most probably cached should use a sorted layout.

\begin{figure}[t]
	\begin{subfigure}[b]{.49\linewidth}
		\centering
    \includegraphics[width=1.\linewidth]{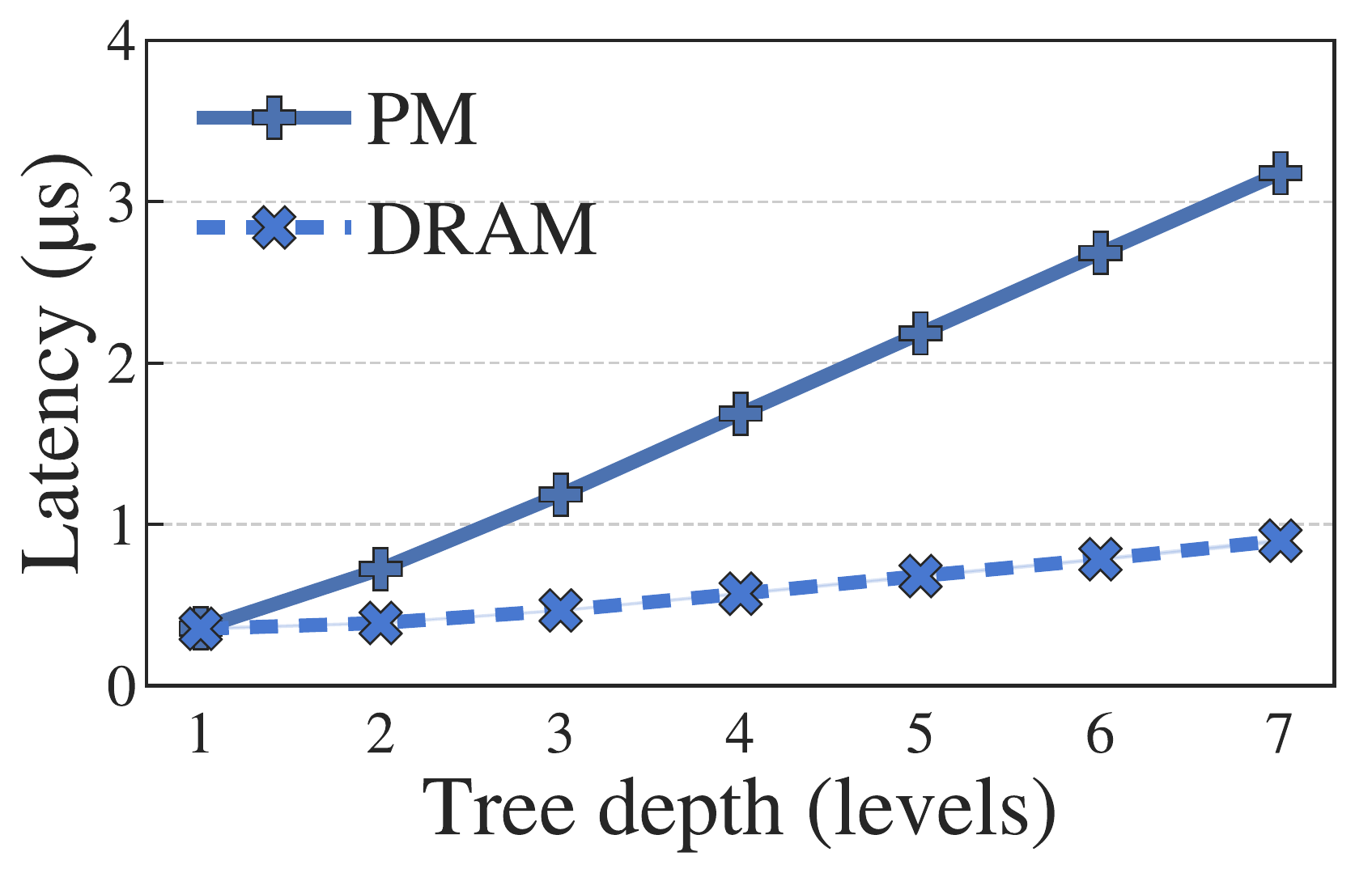}
    \caption{\label{fig:traverse}Traversing a tree w/o search.}
	\end{subfigure}%
  \hfill%
	\begin{subfigure}[b]{.49\linewidth}
		\centering
    \includegraphics[width=1.\linewidth]{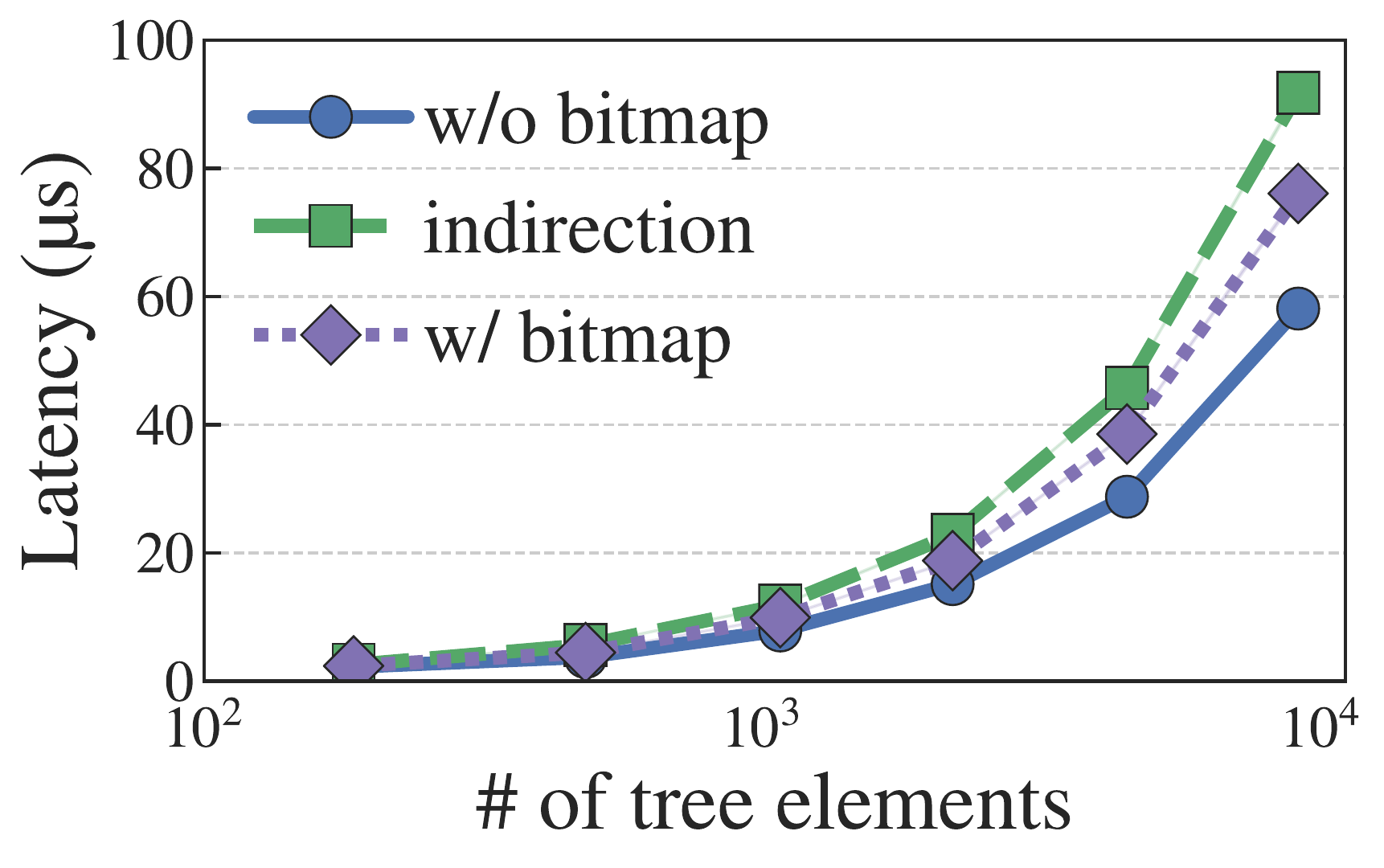}
    \caption{\label{fig:scan}Iterating through nodes.}
	\end{subfigure}
	\caption{\label{fig:traversing}Traversing and iterating nodes (E2 \& E3).}
\end{figure}

\medskip
\textbf{Tree Traversal (E2).}
The next experiment focuses on the inner nodes and the costs for traversing from the root to the leaf level as typical for B$^+$-Trees.
A search within the nodes is not included to get bare dereferencing and pointer chasing measures.
Instead, a random child position is chosen to prevent prefetching.
Therefore, we limit the comparison to the timing of traversing nodes resided in \nvm and DRAM, respectively.
Only the last access is to a persistent leaf node.
This reflects the idea of hybrid data structures and placement. 
Here we have varied the depth of the tree.
The node sizes have hardly made a difference. 
Due to our idle latency measurements for DRAM and \nvm, we would expect an increase of about this latency per level.
The results are shown in \cref{fig:traverse}.

In fact, this behaves almost as expected.
For DRAM, about 50-100~ns are added for each further level.
For \nvm, however, each level adds 400-500~ns, which is nearly double the reported latency of the MLC benchmark.
We assume that this is mainly due to the software overhead (e.g., PMDK) and the random access under heavy load.
However, we also note that this would nearly fit with the reported read latency in~\cite{VLDB:lucas,DBLP:conf/damon/RenenVL0K19}.
It becomes visible that all approaches would greatly benefit from a hybrid variant.
If pure performance in the operating system is most important, we found DRAM-based sorted nodes with both binary search and indirection to be good solutions.
The former is more memory efficient and the latter saves write operations, which however is not so crucial on DRAM.
Placing the inner nodes in DRAM, however, requires recovery actions in the event of a failure.
If this is not desired, it is mainly the search algorithm that makes the difference (see E1).
In summary, if recovery is rare, a hybrid approach is highly preferable.

\medskip
\textbf{Tree Iterate (E3).}
For the last experiment in the read category, the horizontal traversal of data nodes, usually also referred to as scan, is examined.
This contains not only the chasing of the node pointers (like in E2) but also the iteration of the key and value arrays within them.
Since the order is not prescribed, we stick with the term iterate to avoid confusion with range scans.
For this experiment, all approaches use the same number of entries based on the variants with a growing search structure.
Here, we use different data node sizes and let the tree horizontally grow by increasing the single inner node (the root).
Since the order does not matter when iterating, the sorted and unsorted approach use the same algorithm.
The same applies for indirection, hashing, and bitmap, as only the bitmap has to be checked for valid entries.
However, since this causes branching in the loop, we expect a weaker performance of the latter class.
For the indirect organization, it is also possible to iterate using the slot array instead of the bitmap. 
In \cref{fig:scan} the results for 1~KiB data node sizes are reported.

\begin{figure*}
  \centering
  \includegraphics[width=1\linewidth]{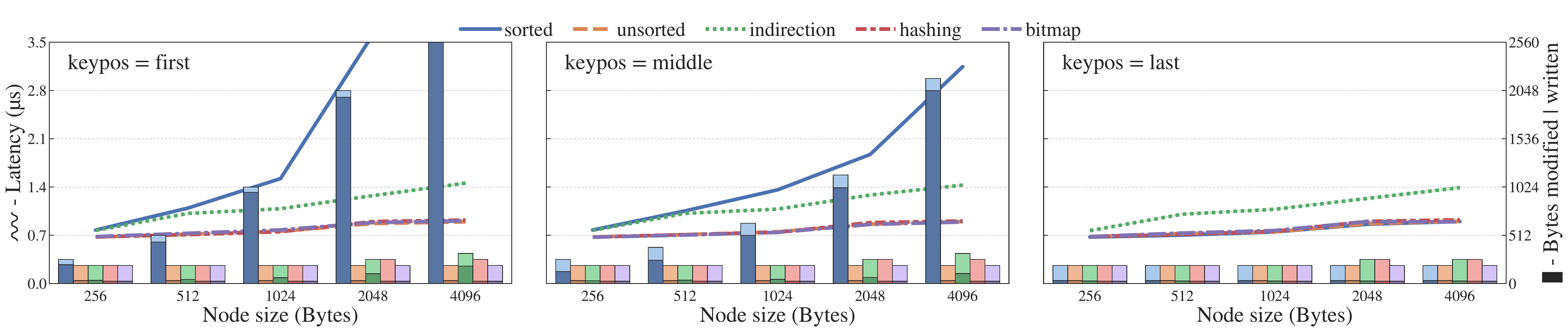}
  \caption{\label{fig:insert}Inserting a key into a node (E4).}
\end{figure*}

Interestingly, iterating via indirection is worse than via bitmap.
This means that even with indirection slots the bitmap should be used for iteration if the order is not important.
As expected the approaches without a bitmap are always the fastest.
Using this as a baseline, in the largest case the overhead is 31\% for the bitmap and 58\% for indirection.
However, it should be noted that the nodes were filled to 100\%.
Thus, it is already the best case for the bitmap since the other approaches also have to check all entries.
In the case of indirection and without a bitmap the loop only iterates through the number of actual keys.
Besides branching, jumps between cache lines (bitmap/indirection slots, key, and value array) also have a negative impact.
Since our scan function only copies out the key in each case, we assume that this is the worst case and that with the increasing complexity of the function all methods might approach each other.
For a range scan based on this, the impact of sorting per node compared to pre-sorted nodes has to be checked.
Nevertheless, for iterating through persistent data nodes the sorted and unsorted approach perform best.
Particularly for the DCPMMs, it is important to avoid jumping between non-sequential cache lines.

\subsection{Insert-based Operations}

\medskip
\textbf{Node Insert (E4).}
For insert-based operations, we first check the behavior when inserting a key-value pair into a data node.
Similar to experiment E1, we vary the node size and the insert position.
In addition to the time an operation takes, we report the number of bytes modified and actually written to the device.
In the setup phase, the key to be inserted is omitted so that space is left for it.
For instance, when inserting at the first position in a node with 10 slots, the keys from 2-10 are pre-inserted.
The insertion of key 1 is then the measured part.
The lookup for the insert position is not part of the measurement.
Adding up the times of traverse, node searches, and node insert would result in approximately the insert macro operation.
We expect that the sorted approach will show the poorest results as entries have to be moved.
It leads to many writes and flushed cache lines.
This is not the case for the other approaches, which only append the new data and adapt the search structure.
The results are illustrated in~\cref{fig:insert}.

It is apparent that the unsorted variants almost always perform best.
For the plain unsorted case, only the key and value are appended and the size field is updated.
The hashing and bitmap approach have to set the corresponding hash and bit, respectively.
Also the indirection performs quite well, even in the first case were all slots have to be shifted to the back to keep the indirect ordering.
Compared to the other approaches optimized for \nvm, the indirection performs worst.
The overhead is not caused by sorting, but by the additional determination of a free bit position beside the given slot position.
Since indirection is the only approach that has to find two positions, we have included this overhead.
Matching the high number of write accesses, the performance of the sorted approach is significantly worse than the others.
It can only keep up with small node sizes.
This is because the number of flushed cache lines is about the same here.
It becomes obvious that keeping the nodes sorted is not suitable for read-write asymmetric \nvm.
Although indirection also involves many writes, these are on a much finer granularity and multiple slots can be persisted at once.
Hence, it shows a similar performance as when using hashing or appending only.
Especially the impact of the read-write asymmetry of \nvm becomes clear by this experiment.
Overall the unsorted variants perform about equally well.

\begin{figure}[h]
 \centering
  \includegraphics[width=.99\linewidth]{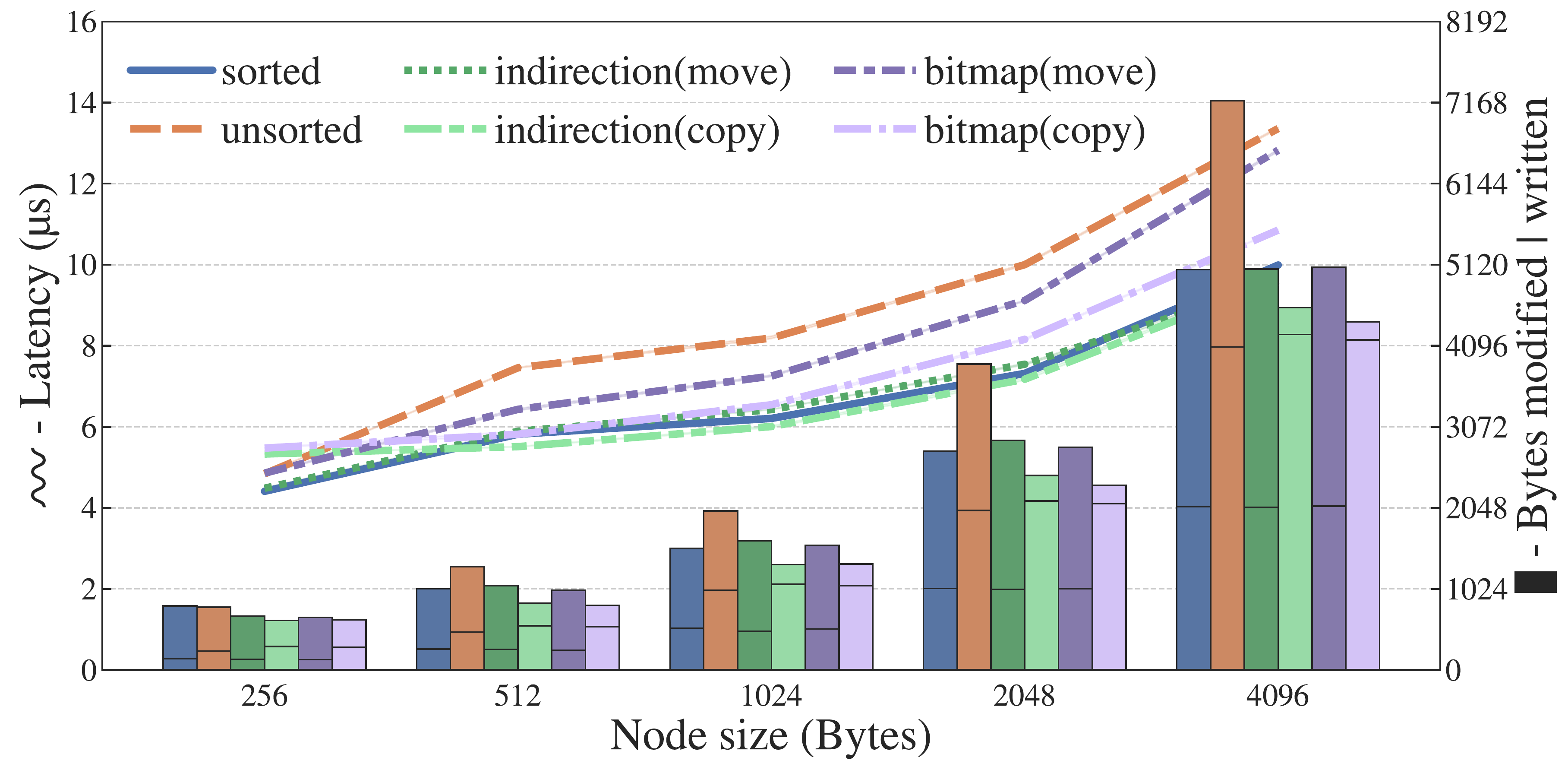}
  \caption{\label{fig:split}Splitting a node (E5).}
\end{figure}

\medskip
\textbf{Node Split (E5).}
As next experiment, we chose node splits in particular for data nodes as these are definitely placed on \nvm.
We picked a similar setup as for the inserts.
We applied the two split strategies as mentioned in \cref{sect:primitives} to the indirection, hashing, and bitmap approach.
Since bitmap and hashing showed exactly the same performance and to keep the figure clear, we summarized them as bitmap in the graph.
As stated before the move variant will cost less writes and thus is supporting DG1, whereas the copy variant exploits the fine-grained access supporting DG2.
For a node organization without bitmap the copy strategy is not useful since the entries of the new node would be written twice.
This is because the whole node is copied and then all entries are reordered to the left.
Generally, the performance is hard to predict for us, but we expect at least that the sorted approach should be faster than the unsorted one.
This is since in an unsorted node all entries have to be checked if they are greater or less the split key.
In a sorted node, everything is simply copied starting from the middle.
\cref{fig:split} shows our results with measures for performance and the number of bytes modified as well as actually written to the device.

In general a split on a sorted node is the fastest and the unsorted case (with or without bitmap) is almost always worst.
Indirection with move strategy behaves similarly to the sorted variant, but requires more effort to transfer and set the indirection slots and bitmap per entry.
Hashing and the bitmap needs more time due to the search for the median in an unsorted array (using quickselect).
The copy strategy needs more bytes as it contains the writing of a whole node.
According to the write endurance and read-write asymmetry, this could be a shortcoming.
However, we combined this copy process with the allocation and as it initiates sequential writing, this seems beneficial for the write-combing buffer on the DCPMMs.
The copy strategy works a bit better since everything is transferred once and after that the slots are shifted and the bitmap is inverted.
The same applies to hashing, so we can deduce that the copy approach is more effective when a bitmap is present.
Also if the inner nodes are in DRAM, we would recommend either a sorted variant or the copy approach.
In this setup, for \nvm most of the time is spend on allocating a new node (around 80\%) and, thus, the approaches show relatively the same performance.
The allocation is done by PMDK encapsulated into a transaction and the time depends on the allocated size.
As already discussed in~\cite{VLDB:lucas} \nvm allocations add a tremendous overhead and should be handled with care.
As suggested in~\cite{DBLP:conf/sigmod/OukidLNWL16} a group allocation could reduce this overhead.
Apart from this, we see a compromise of performance against endurance when having unsorted nodes.
However, also here the sorted and indirection variants are always better.

\begin{figure}
  \centering
  \includegraphics[width=.95\linewidth]{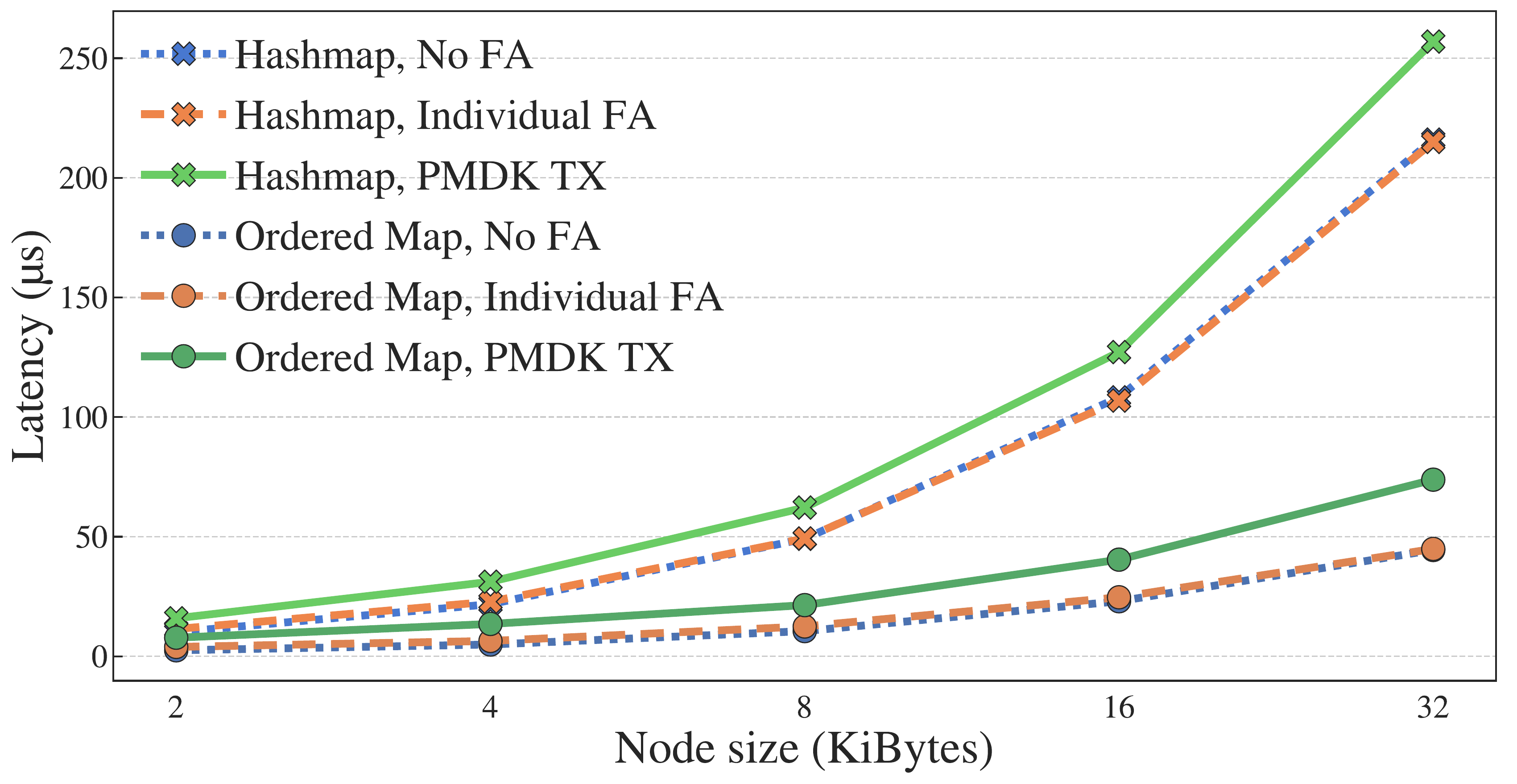}
  \caption{\label{fig:LSM_move}Move data from DRAM to a \nvm node (E6).}
\end{figure}

\medskip
\textbf{Move Node (E6).} In this experiment, we are interested in profiling the latencies involved in moving DRAM data to a persistent node as necessary in LSM-Trees when merging to the first level.
The experimental setup involves varying the node size and switching between the FA strategies: \textit{No FA}, \textit{Individual FA} and the default \textit{PMDK transactions} as explained in~\cref{sect:primitives}.
It is to be noted that all nodes have the same capacity and if the DRAM data is unsorted, then the measurements also involve the sorting operation.
The nodes are composed of persistent arrays where each element is a key-value pair.
We conduct the experiment by inserting unique keys (since this operation is independent of inserts or updates).
The first goal is to analyze the overall \nvm write performance for two cases: (1) Sorting the DRAM data and moving it to \nvm, against (2) Maintaining a sorted DRAM data structure.
The second goal is to analyze the effect of different FA strategies on varying persistent node sizes.
The results are illustrated in~\cref{fig:LSM_move}.

It is apparent that using a sorted DRAM data structure is faster than a unsorted hash data structure since in the unsorted case, each time the capacity is reached, the data must be sorted and moved to a persistent run.
However, maintaining a sorted DRAM data structure is costlier.
For a typical LSM-Tree use case scenario, the DRAM buffer is in the order of a few kilobytes.
Hence, a sorted data structure is always a better choice for small DRAM buffers.
Regarding the second goal, as depicted in~\cref{fig:LSM_move}, using PMDK transactions for FA has a much higher performance impact, when compared to \textit{No FA} and \textit{Individual FA} (cf. \cref{sect:primitives}).
On the other hand, \textit{No FA} and \textit{Individual FA} have almost the same performances, i.e., adding a single 64-bit persistent variable into the PMDK transaction (plus the individual flushes and fences) has negligible performance impact.
This shows that PMDK transactions should only be used for allocations and deallocations.
Performance critical applications should definitely take care of failure atomicity individually.

\begin{figure*}[t]
	\centering
	\includegraphics[width=1\linewidth]{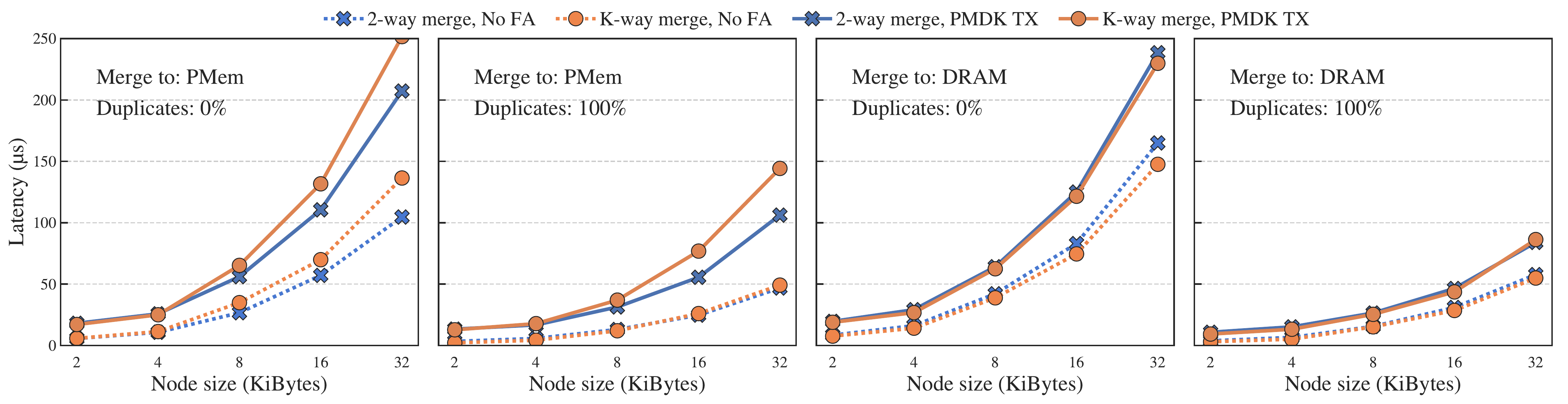}
  \vspace{-2em}
	\caption{\label{fig:LSM_merge}Merging sorted data in persistent nodes to a new persistent node (E7).}
\end{figure*}

\begin{figure*}[t]
  \centering
  \includegraphics[width=1\linewidth]{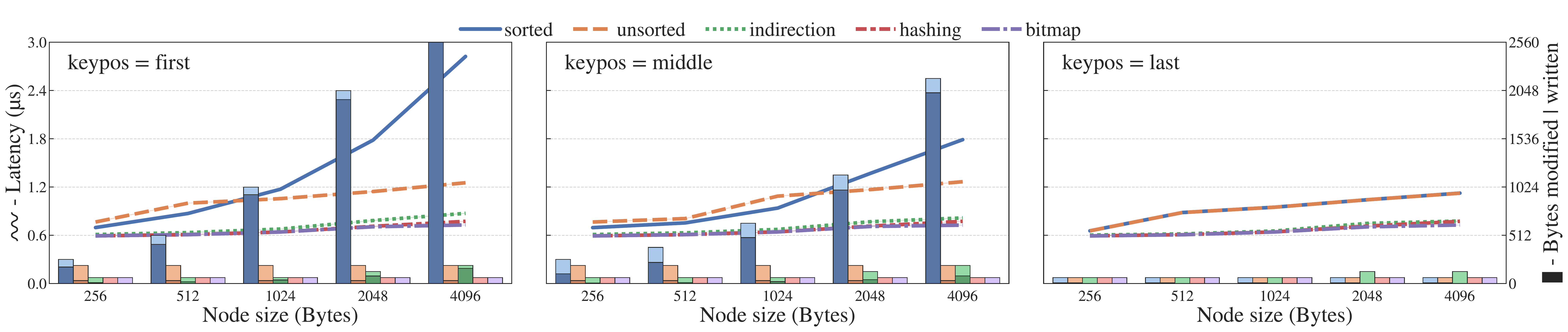}
  \vspace{-2em}
  \caption{\label{fig:erase}Erasing an entry from a node (E8).}
\end{figure*}

\medskip
\textbf{Merge Level (E7).}
In this experiment, we examine the impact of merging sorted \nvm nodes into a new larger node as applied, for example, to the nodes of one level in an LSM-Tree to the next level.
Similar to the previous experiment, the setup involves varying the node size and switching between two different FA strategies (i.e., PMDK transaction and No FA).
Additionally, we examine the impact of the two extreme scenarios: Unique keys in each node and duplicate keys in all the nodes (i.e., 0\%  and 100\% duplicates).
Finally, we benchmark the performance by applying the 2-way and K-way merge algorithms as illustrated in \cref{fig:lsm_2wayMerge} and \cref{fig:lsm_KwayMerge}, respectively.
We used two merge sub-strategies in our experiments.
(1) Merge directly to a \nvm node, (2) Merge to a DRAM buffer and then copy the result to a \nvm node.
These two strategies are applied on both 2-way and K-way algorithms.
The results are shown in~\cref{fig:LSM_merge}

An important observation is that enabling PMDK transactions has again a great performance penalty in all cases.
When merging directly to a persistent node, 2-way is faster since the CPU can cache the intermediate merge results.
On the other hand, in K-way, the CPU needs to read the persistent memory more often for key comparisons due to more cache misses.
It is interesting to see that K-way merge shows the same performance as 2-way when the keys are duplicated in each node (second column of ~\cref{fig:LSM_merge}).
However, when PMDK transactions are enabled the 2-way merge is faster again.
The effect of PMDK transactions is explained as follows.
After a merge operation, the number of elements in the resultant node can vary between the two extreme limits: $\langle$ the number of elements in a single \nvm source node : the sum of elements in all the \nvm source nodes $\rangle$.
In a general scenario, it is not possible to exactly determine the resultant number of elements.
Therefore, in case of K-way merge, the entire sum of elements must be added into PMDK transaction whereas in 2-way merge it is sufficient to add only the sum of elements of the last binary merge step.
Hence, 2-way merge has a better performance when the keys are duplicated with the default PMDK transactions enabled.
To improve the performance of K-way merge, one could use intermediate DRAM buffers as shown in \cref{fig:lsm_2wayMerge} and \cref{fig:lsm_KwayMerge}, respectively.
Once again we see that, there could be two possible scenarios.
(1) No or very few duplicate keys: In such cases, using an intermediate DRAM brings-in additional performance penalty for both the merge algorithms because the number of elements added into a transaction is always equal to the sum of all elements.
(2) Many duplicates: In such cases, both algorithms perform better when an intermediate DRAM buffer is used as shown in column four of ~\cref{fig:LSM_merge}.
The insights from this experiment can be summarized as follows:
(1) 2-way merge could be a preferred choice in the scenarios where the inserted keys are unique and if PMDK transactions are enabled.
(2) K-way merge can be used in the scenarios where no FA is needed and if there are many updates (i.e., duplicates).
(3) If there are many updates and PMDK transactions are enabled, then using an intermediate buffer will result in  performance enhancement for both the merge algorithms.

\subsection{Erase-based Operations}

\medskip
\textbf{Erase from Node (E8).}
For erase-based operations, the first experiment is the removal of a single key-value pair from a node.
Similar to the lookup (E1) and insert (E4), we measure different node sizes and key positions.
Again, we report the latency and modified as well as written bytes for each operation.
The preparation always creates a full node, where the entry is then deleted at different positions.
The lookup for the erase position is again not part of the measurement since it is already represented in E1.
Hence, an erase macro operation without underflow would be the result of traversing the tree, searching each node and this experiment.
Once again, we expect the sorted approach to show the poorest results as keys and values have to be moved to fill the caused gap.
This entails many writes and flushes.
For the unsorted organization, not all the entries have to be shifted.
In this case, it is enough to move the last entry to the caused gap and decrease the entry counter.
The hashing, indirection, and bitmap variants only need to reset one bit.
For indirection the slot array has to be additionally shifted.
Therefore, these approaches will probably run the fastest, with indirection possibly taking slightly more time.
The actual results are visualized in \cref{fig:erase}

The advantage of the bitmap is unambiguously.
In all cases it performs the best.
The hashing approach is directly below the bitmap line, because the algorithm is the same.
As expected, indirection is only a little slower.
Starting from 2~KiB it jumps a bit higher if the key is in the first and middle position.
This is due to the fact that from 2~KiB the bitmap and slots need another cache line to be flushed (cf. ~\cref{tab:keysLeafNode}).
In the last case nothing has to be shifted, thus, it is almost constant.
We would have estimated the unsorted variant to be more stable, since always the same number of bytes are changed.
Here the locality of the deleted and last position, from where the entry is moved, is quite important.
The sorted approach is absolutely not appropriate for erasing a key in \nvm.
It costs way too much writes and also flushes which drastically reduces the performance.
Only when the entry is rather at the end, this approach can keep up.
Hence, using a bitmap is the best choice for fast erasures.

\begin{figure}[b]
  \centering
  \includegraphics[width=1.0\linewidth]{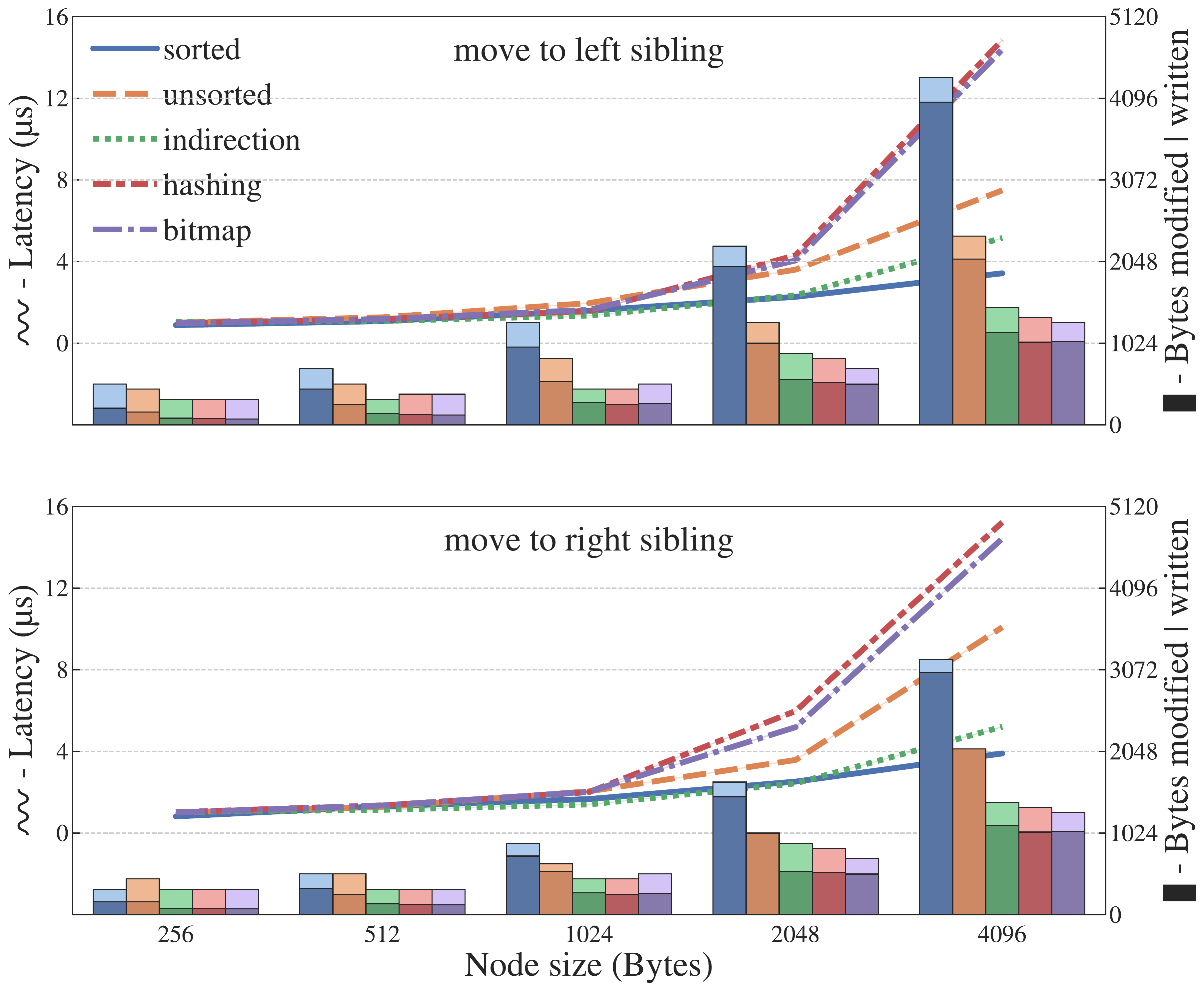}
  \vspace{-2em}
  \caption{\label{fig:balance}Balancing two nodes (E9).}
\end{figure}

\medskip
\textbf{Balance Node (E9).}
Often in trees it is necessary to move entries from one node to another.
For instance, an erase operation can lead to an underflow and a balance operation.
This operation is what we want to evaluate in this experiment.
For this, in the setup, arrays of full nodes and half filled nodes (actually: half-1) are prepared.
The balance operation should move a quarter of the entries in the full node to the half filled node.
There are two possible cases.
Either the entries are moved to a node with smaller or larger keys.
If the order is important the first case requires a shift of the already existing entries on the donor site to bring them to the front.
In the other case, a shift is necessary on the receiver site to make place for the new smaller entries.
Since the number of writes is about the same, we do not expect much difference.
Again, we tested for various node sizes and report the average latency as well as the modified bytes.
Basically, we expect the sorted variants to perform better than the unsorted ones.
The results are shown in \cref{fig:balance}

Unlike our previous experience, the number of written bytes is not reflected in the performance here.
The sorted variants are, as expected, faster but the distance to the other techniques is enormous for larger node sizes.
Comparing the sorted and hashing approach the former is nearly four times faster than the latter.
This is because in the unsorted case the next maximum or minimum must be searched before each move.
In addition, the bitmap approaches must always search and set a free bit on the receiver site.
The hashing approach must also copy the hashes, which can lead to further written cache lines.
Compared to the sorted case, with indirection the slots have to be additionally shifted and written.
However, the keys and values can simply be appended.
Since less is written here, we would have expected it better as direct sorting.
This is only the case with small node sizes.
Nevertheless, with regard to our design goals, we would still consider indirection as the best choice here.

\begin{figure}[t]
  \centering
  \includegraphics[width=1.0\linewidth]{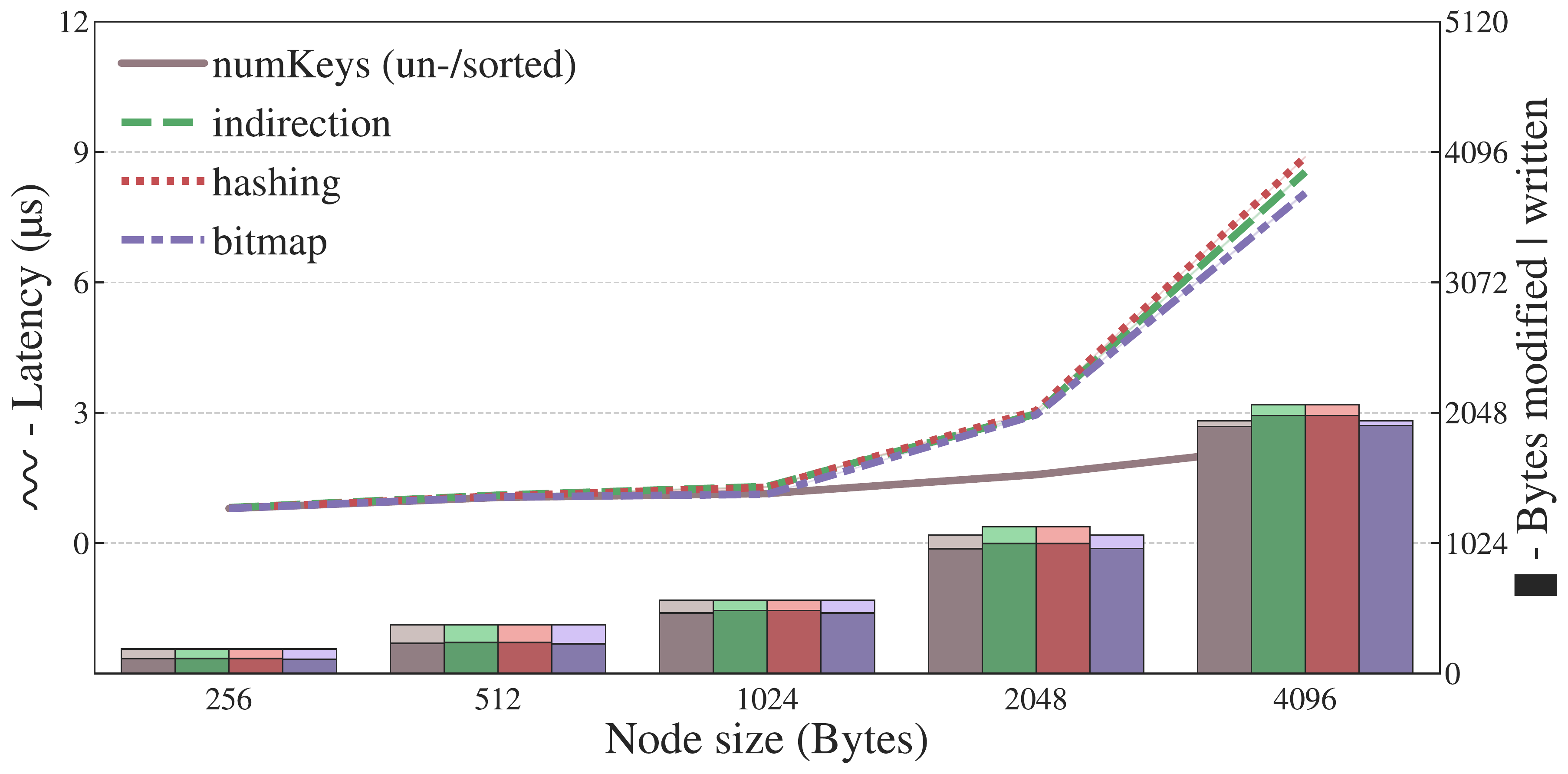}
  \vspace{-2em}
  \caption{\label{fig:merge}Merging two nodes (E10).}
\end{figure}

\medskip
\textbf{Merge Nodes (E10).}
We already discussed the merge of multiple nodes into another in experiment E7.
There is also a less complex merge operation as found as a consequence of an underflow in a B-Tree.
Here, no duplicates are present in this micro-operation.
On top of that, we cannot apply different merge strategies like 2-way or K-way merge since only two nodes are affected.
Instead the various node organizations and their corresponding access primitives are compared once again.
In contrast to the previous experiment, we only consider one direction, the merging into the node with the smaller keys.
This is always the better option, because there is no need to shift entries or slots.
As a result, the sorted and unsorted approach proceed exactly the same.
They simply append all keys and values and finally update the number of keys.
Hence, they are summarized as \texttt{numKeys}.
The deallocation of the donor node is not part of the measurement, but it would be the same overhead for all approaches.
Once more, we report the latency and modified/written bytes.
We expect that the numKeys approach performs better than the approaches using extra search structures since they do not include a mechanism yielding into any performance gain.
In \cref{fig:merge} the results can be inspected.

\begin{figure*}[t]
  \centering
  \includegraphics[width=\linewidth]{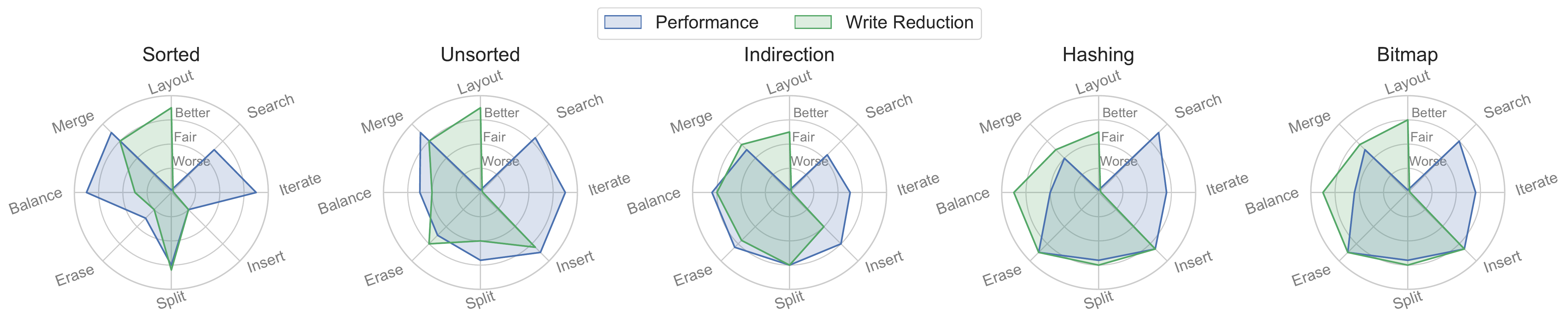}
  \caption{\label{fig:radar}Performance profile of design primitives.}
\end{figure*}

Again the disadvantage of the bitmap shows up.
It requires the verification of each bit on the donor site and search for a free bit on the receiver site before moving an entry.
For indirection the first check is not required, however additional reads are necessary due to the indirect access and the updates of the slot array.
This results in bitmap and indirection having almost the same performance.
Again the hashing approach has to additionally copy the hashes and check the bit on the donor site.
This leads to more written cache lines with larger node size.
Comparing the numKeys and hashing approach, we can again see a performance difference of a factor of four.
In a separate measurement the pure deallocation took constantly around $1.6~\mu s$ independent of the size.
In total, the sorted and plain unsorted variant work most efficiently when merging two nodes.

\subsection{Performance Profiles}
Using the results from our experiments, we created performance profiles summarizing the performance and write reduction of the main identified primitives in this paper (see \cref{fig:radar}).
This bases on the node layout (sorted, unsorted, indirection, hashing, and bitmap) and their corresponding access primitives, from which we considered the best performing alternatives.

For \textbf{sorted} nodes, the main drawbacks are raw entry inserts and key deletions.
Also, the search performance is worse than the other variants in uncached cases.
However, it performs well in iterations and structural adjustments.
Although it is the smallest in size, it takes the most writes and flushes when modifying nodes.

The \textbf{unsorted} layout significantly improves the most typical operations like search, insert and erase, particularly in terms of memory efficiency and write reduction.
If many scans are used, the sorted and unsorted variants are best suited.
Only balancing and splitting costs considerably more as the order is beneficial for these micro-operations.
So if the tree grows and shrinks a lot, this can lead to enormous overhead.
A countermeasure could be a larger node size to avoid too much restructuring.

Another variant to overcome this is to add \textbf{indirection} slots which deliver almost the same performance as in the sorted case, but require less write operations.
This is paid for with a node size overhead and worse read operations.
Therefore, this variant seems more suitable for write dominated workloads.

The \textbf{hashing} approach is a bit worse regarding the restructuring.
However, it offers the best overall package, especially with a small node size.
Similar to the simple unsorted approach the basic micro-operations search, insert, and erase of a key or entry are its great strength.
A drawback is the iterate operation due to the bitmap.
If no underflow handling and few scans are required, this primitive is the best choice.

The bare \textbf{bitmap} approach behaves similarly to the hashing approach and is only slower for a few operations.
Therefore, the combination of hash and bitmap should usually be chosen. 
The sole advantage of the bitmap-only variant is the lower memory consumption and to an extent faster underflow operations.


\section{Insights \& Conclusion}
\label{sect:conclusion}

The results of the experiments gave us some interesting general insights for designing data structures, choosing corresponding access primitives, and combining various ideas.
In part, the insights gained were consistent with these in~\cite{VLDB:lucas}.

\circled{I1:} As it became clear already from looking at the design space, there are still numerous untested possible primitives and combinations of them.
For instance, using hash probing without a bitmap but a \emph{numKeys} field, combining indirection for inner and hashing for data nodes, examining other algorithms like interpolation or exponential search, etc.
Also investigating well-known techniques like compression or zone maps to reduce writes and read accesses seems reasonable.
Explicitly combining the tested primitives for B$^+$-Trees, we propose hash probing and bitmap for data nodes (1~KiB) and a sorted layout for inner nodes residing in DRAM.
If inner nodes also need to be persistent and there are many writes, they should rely on indirection.
This is basically a combination of the ideas from~\cite{DBLP:journals/pvldb/ChenJ15} and~\cite{DBLP:conf/sigmod/OukidLNWL16}.

\circled{I2:} A hybrid DRAM/\nvm approach is highly recommended when seeking the best performance and still requiring persistence.
Especially the traversal experiment E2 proved that dereferencing and pointer chasing has an even greater impact on \nvm.
In our DRAM-based and cached tests, we found sorted and indirection to be the best solutions for inner nodes.
If a hybrid approach or recovery is not an option, it is important to note that hash probing is not applicable for inner nodes.

\circled{I3:} PMDK transactions are universal, but not recommended for performance critical applications.
As it was evident in E6 and E7 the log and snapshotting used in PMDK transactions add a tremendous overhead compared to individual realizations of failure atomicity.
This means that the classic Copy-on-Write approaches should be avoided for \nvm.

\circled{I4:} Jumping between non-sequential cache lines is quite expensive.
Although \nvm allows byte-addressable random access, sequential access is still preferable.
This was particularly apparent for indirection and binary search (E1), iterations via bitmap and indirection (E3), as well as erasing in unsorted nodes (E8).
Especially the latter showed the importance of locality.

\circled{I5:} Allocations are expensive in \nvm and depend on the requested size.
During the experiments E5 and E10 it became clear that allocations should be used wisely.
To overcome this bottleneck, designers of \nvm-based data structures should use group allocations and reuse already allocated nodes instead of frequent deallocating and allocating.

\circled{I6:} The optimal size for nodes located in \nvm-based index structures lies between 256 bytes and 1 KiB.
The lower bound of 256 bytes results from the write-combing buffer of the DCPMMs.
The upper bound is the size from which the performance typically drastically degrades.
This is partly due to the search structures, which should not grow beyond a cache line.
Small nodes automatically lead to longer traversing routes, hence we refer back to I2.

Ultimately every design decision depends on the specific application and there is no single best solution.
This paper is intended to help determine the optimal \nvm-specific design parameters and (combinations of) primitives for a given use case.
For instance, for a write-intensive workload with many structural adjustments, indirection is the primitive of choice.
If there are many point queries, hashing is best suited.
To accelerate deletes, a bitmap should be added.
If mainly iteration through the data nodes is required, no auxiliary structure should be used.
\Cref{tab:primitives} and our investigations still offer much potential for extension.
For future work, further data structures, primitives, and access patterns shall be studied.
The final result should be a far broader derived performance profile per design primitive as sketched in~\cref{fig:radar}.

\begin{acks}
This work was funded by the German Research Foundation (DFG) within the SPP2037 under grant no. SA 782/28.
\end{acks}

\bibliographystyle{ACM-Reference-Format}
\bibliography{main}


\end{document}